\documentclass{ametsocV6.1}

\title{Data-driven Seasonal Climate Predictions via Variational Inference \\ and Transformers}

\authors{Lluís Palma,\aff{a,b}\correspondingauthor{Lluís Palma, lluis.palma@bsc.es} 
Alejandro Peraza,\aff{a}
David Civantos,\aff{c} 
Amanda Duarte,\aff{a} 
Stefano Materia, \aff{a} \\
Ángel G. Muñoz, \aff{a}
Jesús Peña-Izquierdo,\aff{c}
Laia Romero,\aff{c}
Albert Soret\aff{a}
and Markus G. Donat\aff{a,d} 
}

\affiliation{\aff{a}{Barcelona Supercomputing Center, Earth Sciences Department, Spain }\\
\aff{b}{Facultat de Física, Universitat de Barcelona, Martí i Franquès 1, 08028 Barcelona}\\
\aff{c}{Lobelia Earth, Barcelona, 08005, Spain }\\
\aff{d}{Institució Catalana de Recerca i Estudis Avançats (ICREA), Barcelona, Spain}\\
}

\abstract{Most operational climate services providers base their seasonal predictions on initialised general circulation models (GCMs) or statistical techniques that fit past observations. GCMs require substantial computational resources, which limits their capacity. In contrast, statistical methods often lack robustness due to short historical records. Recent works propose machine learning methods trained on climate model output, leveraging larger sample sizes and simulated scenarios. Yet, many of these studies focus on prediction tasks that might be restricted in spatial extent or temporal coverage, opening a gap with existing operational predictions. Thus, the present study evaluates the effectiveness of a methodology that combines variational inference with transformer models to predict fields of seasonal anomalies. The predictions cover all four seasons and are initialised one month before the start of each season. The model was trained on climate model output from CMIP6 and tested using ERA5 reanalysis data. We analyse the method's performance in predicting interannual anomalies beyond the climate change-induced trend. We also test the proposed methodology in a regional context with a use case focused on Europe. While climate change trends dominate the skill of temperature predictions, the method presents additional skill over the climatological forecast in regions influenced by known teleconnections. We reach similar conclusions based on the validation of precipitation predictions. Despite underperforming SEAS5 in most tropics, our model offers added value in numerous extratropical inland regions. This work demonstrates the effectiveness of training generative models on climate model output for seasonal predictions, providing skilful predictions beyond the induced climate change trend at time scales and lead times relevant for user applications.
} 

\begin{document}

%% Necessary!
\maketitle

\section*{Introduction}\label{sec:intro}

In contrast to weather forecasts, which predict daily atmospheric conditions for up to two weeks, seasonal climate predictions provide estimates of seasonal statistics months in advance. To address the inherent unpredictability of the atmosphere \citep{lorenzDeterministicNonperiodicFlow1963} and the resulting stochasticity of the Earth system, seasonal climate predictions leverage ocean and land surface forcings in conjunction with ensemble predictions that quantify uncertainty \citep{palmerProspectsSeasonalForecasting1994}. Thus, seasonal outlooks commonly deliver probabilities of wetter/drier or warmer/colder than average conditions. This information has proven valuable for many climate-sensitive sectors, including agriculture \citep{challinorProbabilisticSimulationsCrop2005, perez-zanonLessonsLearnedCodevelopment2024}, renewable energy production \citep{garcia-moralesForecastingPrecipitationHydroelectric2007, lledoSeasonalPredictionEuroAtlantic2020, ramonPerfectPrognosisDownscaling2021} or public health \citep{thomsonMalariaEarlyWarnings2006}. Consequently, over the past few decades, seasonal climate prediction has transformed from a research effort into an operational service \citep{sahaNCEPClimateForecast2014, Johnson2019}. Nonetheless, the practical value of seasonal predictions relies on their skill and resolution, and for many applications, those might not reach user requirements \citep{Doblas-Reyes2013a}.\\

Current operational climate services providers base their seasonal predictions on dynamical models, statistical methods, or a hybrid combination. Dynamical models are based on coupled (atmosphere, land, ocean, and sea-ice) General Circulation Models (GCMs), which embody the most complete representation of climate system dynamics known. These models are typically initialised to our best estimate of the observed climate state \citep{meehlInitializedEarthSystem2021}, integrating the diverse mix of observations available into a consistent set of fields through a process known as data assimilation. Those fields provide an initial state of the Earth system and serve as a starting point for a simulation covering the desired prediction period, typically up to eight months for seasonal predictions. Simulating the Earth system at a global scale over long periods and multiple ensemble members requires vast computational resources, limiting the model's spatial resolution \citep{scaife2019atmores}. Such limitation cascades into many physical processes not being explicitly resolved and subject to parametrization, resulting in biased dynamics. These biases add to an imperfect definition of the initial state, partly due to an erratic spatial and temporal distribution of current observations \citep{materia2014impact, ardilouzeMultimodelAssessmentImpact2017a}. This combination of factors results in seasonal predictions presenting strong drifts and biases, deteriorating prediction quality \citep{Doblas-Reyes2013a, weisheimerReliabilitySeasonalClimate2014}.\\

On the other hand, statistical or empirical prediction methods benefit from efficiently leveraging the relationships learned from past observational records. They explicitly capture the interactions between predictors and predictands, offering a more direct approach than dynamical models, which derive such relationships through iterative simulation at finer temporal scales. Statistical methods range from simple persistence models to sophisticated statistical techniques \citep{edenGlobalEmpiricalSystem2015}, including machine or deep learning algorithms \citep{haoSeasonalDroughtPrediction2018}. These approaches can yield skillful forecasts comparable to dynamical models. Still, statistical models are not exempt from errors and require careful application due to short observational records and climate non-stationarity, which often compromises the independent and identically distributed (i.i.d.) assumption, paramount for many of these methods.\\

The limited temporal extent of current observational datasets poses a greater challenge when involving large machine learning (ML) algorithms. Training these algorithms with a limited dataset results in almost certain overfitting due to the imbalance between trainable parameters and available training samples. Unlike weather forecasting, where high-frequency temporal variability yields multiple independent samples over short periods, seasonal processes operate on monthly to annual scales. At these timescales, interannual processes dominate, resulting in as few as one independent sample per year. Consequently, seasonal forecasting applications have far fewer data points, up to two orders of magnitude less than weather applications trained over the same period \citep{anderssonSeasonalArcticSea2021, gibsonTrainingMachineLearning2021, materiaArtificialIntelligenceClimate2024}. This limitation partly explains the explosion in studies applying large deep learning models to weather forecasting \citep{biPanguWeather3DHighResolution2022, pathakFourCastNetGlobalDatadriven2022, lamLearningSkillfulMediumrange2023, priceGenCastDiffusionbasedEnsemble2024, kochkovNeuralGeneralCirculation2024}, contrasting with the few works tackling seasonal climate predictions.\\

To address such limitations, most ML applications for seasonal predictions rely on climate model output to train their data-driven models \citep{hamDeepLearningMultiyear2019, gibsonTrainingMachineLearning2021, felscheApplyingMachineLearning2021}. The underlying assumption is that climate models can, to a certain extent, simulate the climate system and its interannual variability, making the thousands of simulated years a valid training set. Beyond deep learning, this logic has also motivated works in the climate community where analogues from climate model output assemble seasonal to decadal predictions from initial conditions \citep{dingSkillfulClimateForecasts2018, mahmoodConstrainingLowfrequencyVariability2022,cosNearTermMediterraneanSummer2024, donatImprovingForecastQuality2024}. Some even employ deep learning to select the best analogues \citep{raderOptimizingSeasonalDecadalAnalog2023}. Training with climate model output effectively increases the number of samples from tens to thousands when using multiple simulations. This approach also delivers data from various multi-decadal periods, as the simulations cover hundreds of years, and allows learning from unobserved regime shifts or trends, such as the one forced by global warming. Data that captures regime shifts and unseen scenarios is crucial for most ML algorithms, which typically struggle with extrapolation and are highly sensitivity to regime shifts due to their reliance on the independent and identically distributed (i.i.d.) assumption \citep{belkinReconcilingModernMachine2019, curthClassicalStatisticalInSample2024a}. However, as previously mentioned, climate models (GCMs) have known drifts and biases and misrepresent or fail to resolve critical physical processes, resulting in poor simulation of some key teleconnections that contribute to predictability. Consequently, ML models trained on climate model output are potentially limited by the climate model's ability to simulate the relevant processes affecting the particular seasonal prediction task of interest.\\

\begin{figure}[t]
 \centerline{\includegraphics[width=39pc]{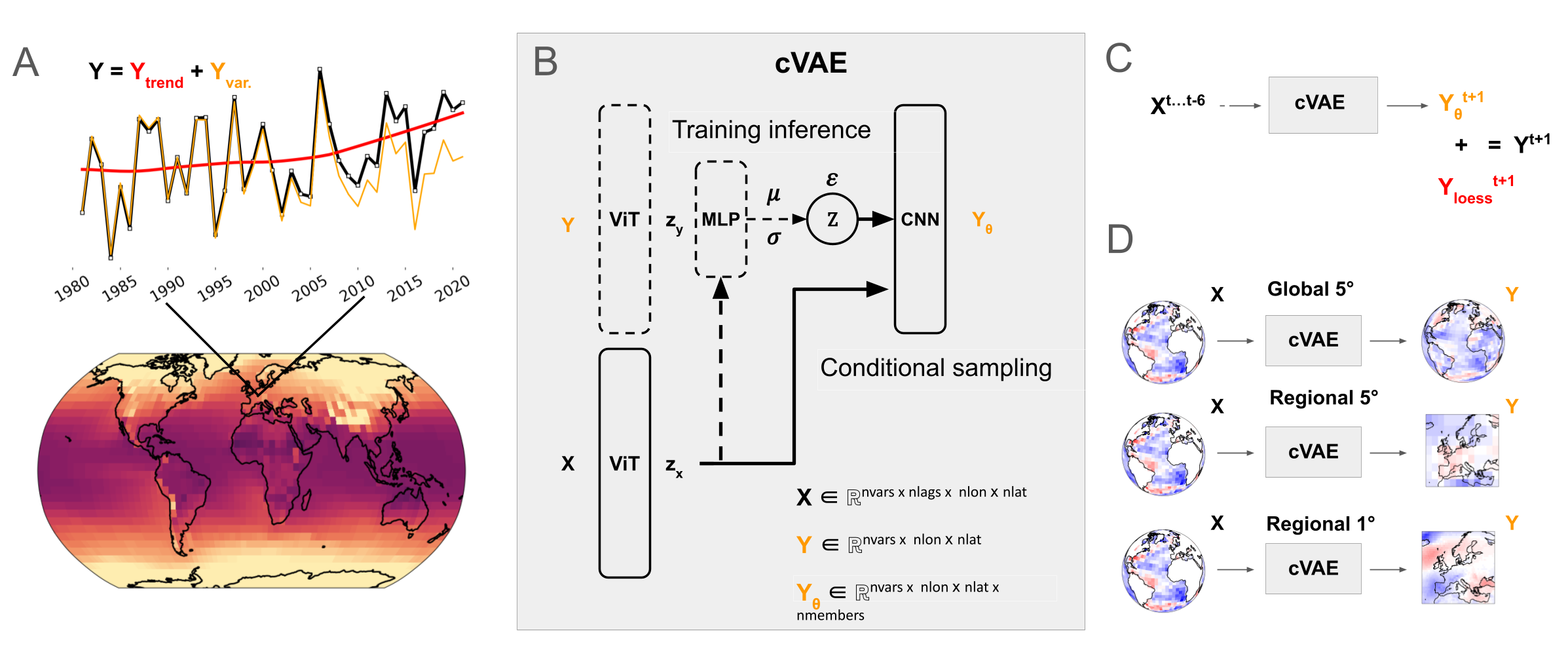}}
  \caption{\textbf{Methodology overview.} \textbf{A}, Illustration of the signal decomposition of the target variable Y. \textbf{B}, Schematic representation of the conditional Variational Autoencoder (cVAE) architecture. Two vision transformers (ViTs) encode the information from multiple climate fields into the latent space. The compressed latent space representation is then passed to the CNN decoder that reconstructs the predicted climate fields. \textbf{C}, Final model assembling, combining the interannual variability prediction from the cVAE model and the regressed LOESS trend. \textbf{D}, Tested model configurations, combining different spatial resolutions and target domains.}\label{f1}
\end{figure}

Current applications of deep learning algorithms trained on climate model output include the prediction of El Niño-Southern Oscillation (ENSO) and the Indian Ocean Dipole (IOD) \citep{hamDeepLearningMultiyear2019, lingMultitaskMachineLearning2022}, the prediction of Arctic sea-ice \citep{anderssonSeasonalArcticSea2021}, the occurrence of drought events in the US \citep{gibsonTrainingMachineLearning2021},  or the prediction of European summer heatwaves \citep{beobide-arsuagaSpringRegionalSea2023}  and droughts \citep{felscheApplyingMachineLearning2021}. Most studies rely on classical machine learning methods or simpler deterministic neural networks and focus on specific prediction tasks that tend to collapse spatial information for simplicity. Providing global probabilistic predictions with data-driven models is essential for a robust comparison against current state-of-the-art dynamical prediction systems \citep{edenGlobalEmpiricalSystem2015}. As an example, \citep{panImprovingSeasonalForecast2022} uses a probabilistic method to predict (Oct-Mar) precipitation and temperature anomalies based on the previous July's upper ocean thermal status. However, how skill varies across seasons remains to be explored, especially at lead times closer to the initial state, which is of particular interest to users. In addition to that, in the context of global warming where strong trends provide added predictability on top of interannual fluctuations \citep{pattersonStrongRoleExternal2022,prodhomme2021seasonal}, it is essential to disentangle both sources of predictability to properly assess the added value of the tested prediction systems \citep{Doblas-Reyes2013a, goddardVerificationFrameworkInterannualdecadal2013, tippettTrendsSkillSources2024}.\\

This study evaluates the effectiveness of an intrinsically probabilistic deep learning method for predicting global three-month seasonal anomalies of temperature and precipitation fields throughout the year. The model is trained using the output of CMIP6 and validated against ERA5 reanalysis data. It explicitly decomposes the contribution of climate change-induced trends during both training and validation. Thus, the approach combines variational inference with vision transformers to explicitly predict interannual seasonal anomalies. To the best of our knowledge, this is the first application of vision transformers for seasonal prediction. Additionally, we apply this methodology in a regional context, specifically focusing on Europe, where we compare the effects and robustness of targeting different spatial domains and resolutions.

\section*{Results}\label{sec:results}

\subsection*{Preliminary performance assessment}

The presented approach (illustrated in Figure \ref{f1}) uses a conditional Variational Autoencoder (cVAE; \cite{kingmaAutoEncodingVariationalBayes2013,kingmaIntroductionVariationalAutoencoders2019}) architecture to predict seasonal climate anomalies. The model takes monthly means of five essential climate variables from the preceding 6 months as input: 2-meter air temperature (tas), precipitation (pr), sea surface temperature (tos), and geopotential height at 500hPa and 300hPa levels (zg500, zg300), and predicts their 3-month seasonal averages. Two vision transformers \citep{dosovitskiyImageWorth16x162021} encode the input and target states into a latent space, capturing the underlying climate patterns. 
Transformers are based on a general-purpose inductive bias that separates the interaction range from the network's depth. This separation enables the modeling of both distant and local connections without requiring a complex hierarchy of convolutional neural network (CNN) operations. As a result, Vision Transformers (ViTs) can more explicitly capture both local and long-range climate interactions. To decode and generate the probabilistic predictions, we employ a CNN that processes multiple samples from the latent space to create an ensemble forecast. The model explicitly predicts the interannual variability component, while a separate locally estimated scatterplot smoothing (LOESS) regression handles the long-term trend. The output of these components is later combined to produce the final forecast. This setup represents an example of using simple statistical methods to capture predictable signals while leveraging neural networks to model complex deviations. The model was trained using Coupled Model Intercomparison Project Phase 6 (CMIP6) data and tested against the European Centre for Medium-Range Weather Forecasts Reanalysis v5 (ERA5) dataset. Further details about the methodology can be found in the Methods section.\\

Seasonal climate predictions are influenced by multiple processes operating at different time scales with varying degrees of predictability. Predictability at the seasonal time scale is affected not only by interannual fluctuations but also by longer-term modulations such as trends driven by greenhouse gas emissions or lower-frequency decadal oscillations. These longer-term processes may have different levels of predictability at the interannual scale. However, seasonal predictions aim to provide information on seasonal anomalies at the interannual level, going beyond trends or decadal oscillations. Thus, disentangling all these different signals in our validation procedure is crucial to understanding the value of the seasonal predictions, if any.\\

\begin{figure}[!h]
 \centerline{\includegraphics[width=39pc]{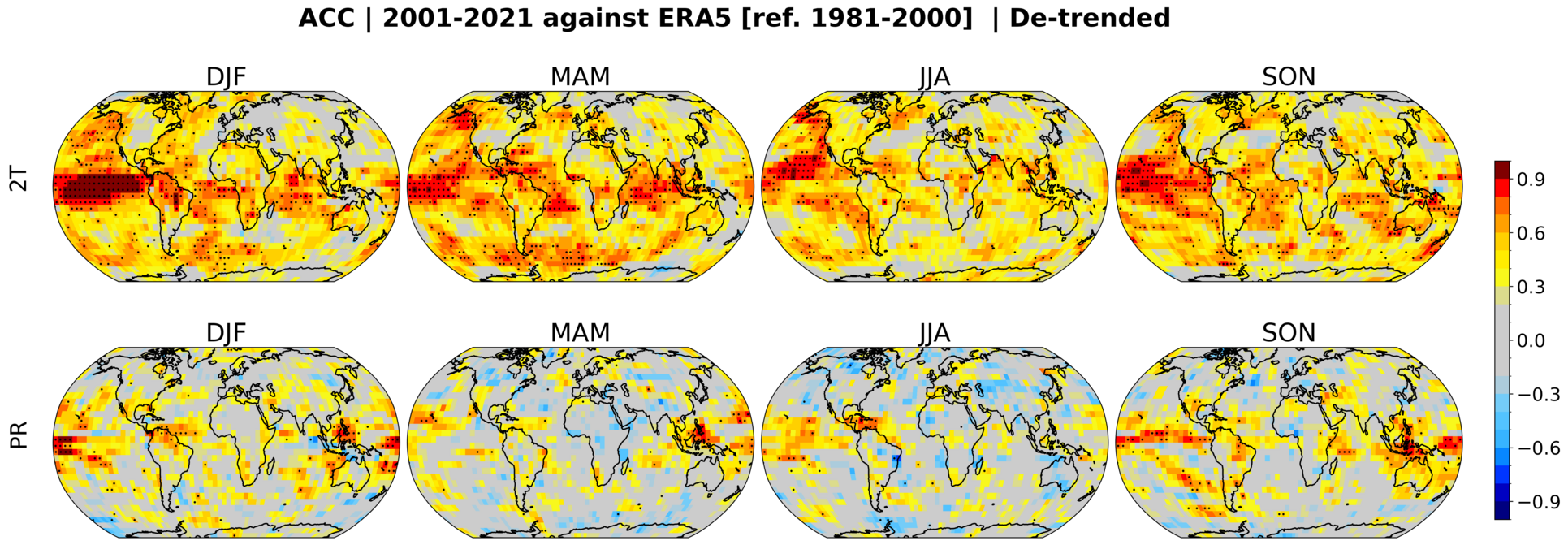}}
  \caption{ACC against ERA5 reanalysis (2001-2021), for temperature (2T, top row) and precipitation rate (PR, bottom row). Seasons are shown in the columns: DJF (December-January-February), MAM (March-April-May), JJA (June-July-August), and SON (September-October-November). Black dots indicate statistical significance at the 95\% confidence level.}\label{f_validation}
\end{figure}

Figure \ref{f_validation} shows the Anomaly Correlation Coefficient (ACC), computed between our predictions and ERA5 reanalysis for both de-trended temperature (trended results are shown in Figure \ref{f_A1}) and precipitation seasonal anomalies. The predictions were initialized one month before the start of the season, i.e. the prediction of DJF used climate information up to November 1st. The ACC validation against ERA5 reveals a generally higher skill for temperature predictions compared to precipitation, with most of the skill concentrated in the tropics and limited skill in the extratropics, with few exceptions such as the west coast of North America, North Atlantic, the Arctic Sea or some parts of Europe and Eurasia.\\

Temperature forecasts exhibit stronger correlations over oceanic regions, with particularly robust signals in the equatorial Pacific across seasons, peaking in the SON and DJF seasons. Beyond the ENSO signature in the equatorial Pacific is the Pacific Decal Oscillation signature that extends into the extra-tropical north-Pacific sea. Significant correlations in temperature are observed over land areas in Central America, Brazil, Australia, central and north (DJF) Africa, South Africa (DJF), and southeast central Asia. Although more limited, Europe also shows positive correlations, especially during the MAM and SON seasons. Precipitation forecasts, while generally less skillful and less spatially extent than temperature predictions, show high correlations (above 0.7) in Indonesia (SON, DJF, MAM) and the Caribbean (JJA), with moderate correlations (0.5-0.7) in northern South America (SON, DJF), Australia (SON), the Horn of Africa (SON), and the US (SON). Weaker (0.3-0.5) precipitation correlations are observed in similar regions for other seasons, including parts of Europe, India, the central Atlantic coast of Africa, and southern South America. Results for an extended period 1985-2021 are shown in the Supplementary material (Figure \ref{f_glob_acc_det_extended}). These results are not directly assessed in the main manuscript as the time period employed overlaps with the reference period defined to compute the standardization and de-trending \citep{risbeyStandardAssessmentsClimate2021a}. Overall, our method captures predictable signals that are physically realistic and consistent with findings reported in the literature \citep{Doblas-Reyes2013a, edenGlobalEmpiricalSystem2015}, demonstrating the potential of the methodology for seasonal forecasting applications.\\

\begin{figure}[h]
 \centerline{\includegraphics[width=29pc]{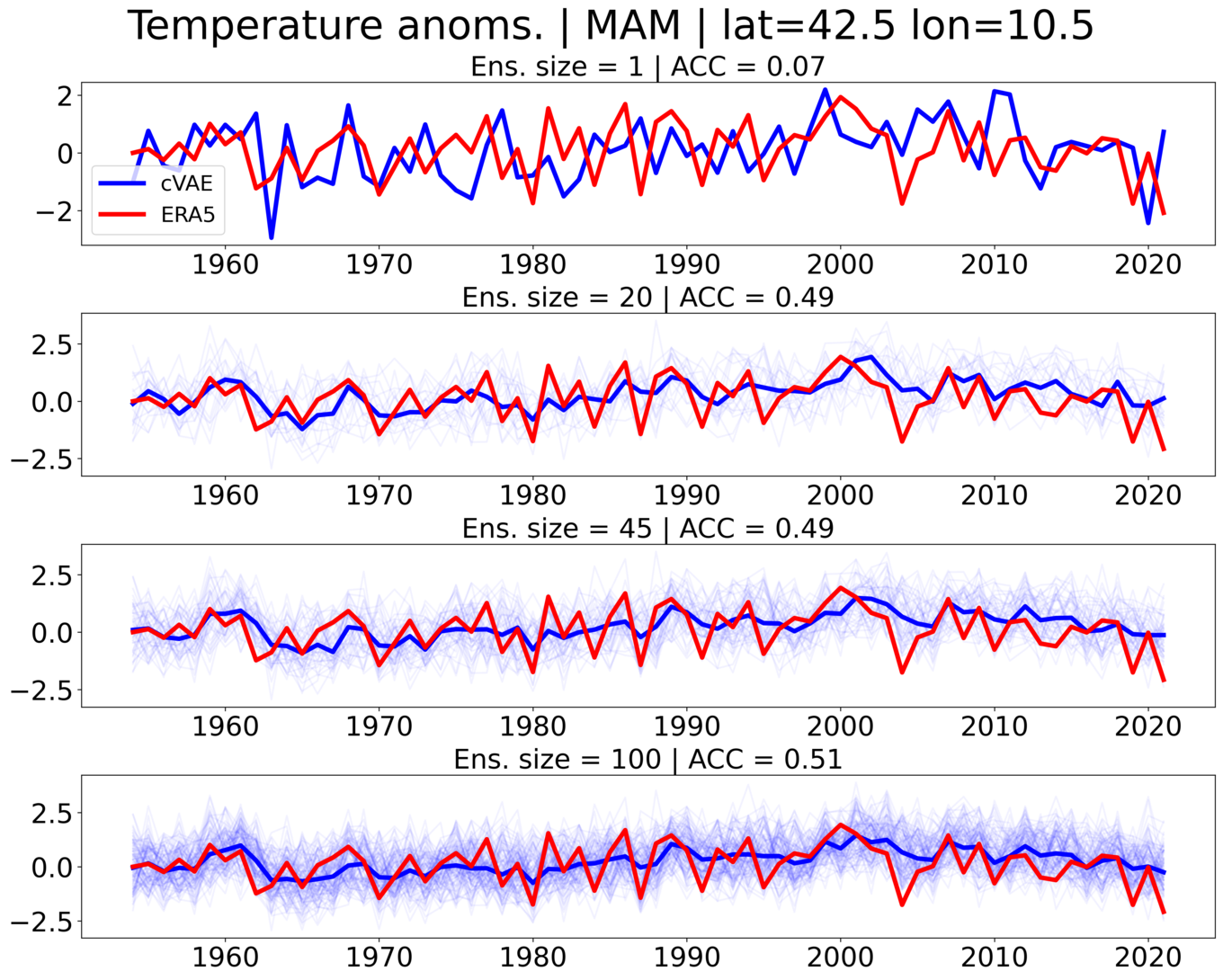}}
  \caption{\textbf{Ensemble size effect on model's deterministic performance.} Effect of ensemble size on the ACC of the ensemble median (blue thick line) against ERA5 (red thick line). The figure depicts MAM temperature anomalies in a grid cell located at 42.5N 10.5E. Blue thinner lines depict the individual ensembles.
}\label{f_ensemble}
\end{figure}

The generative nature of variational methods (described in the Methods section) enables the production of multiple deterministic predictions, forming an ensemble of plausible outcomes. This ensemble facilitates the derivation of a more robust deterministic signal. We make an initial assessment of the ensemble and the impact of its size on the model's performance, shown in Figure \ref{f_ensemble}. We observe that increasing the ensemble size from 1 to 150 members enhances forecast skill, with the ACC improving from 0.07 to 0.51. The model reaches near-optimal performance with an ACC of 0.49 using just 20 ensemble members, indicating a balance between computational efficiency and forecast accuracy. Larger ensemble sizes provide a more explicit representation of forecast uncertainty, while individual members reflect realistic temporal variability, closely following observed climate variability. The model effectively captures key aspects of ensemble forecasting systems while offering potential computational benefits. The ability of generating diverse ensembles while obtaining peak correlations with moderate ensemble sizes are desirable properties for climate prediction.

\subsection*{Validation against benchmarks}

\begin{figure}[!h]
 \centerline{\includegraphics[width=39pc]{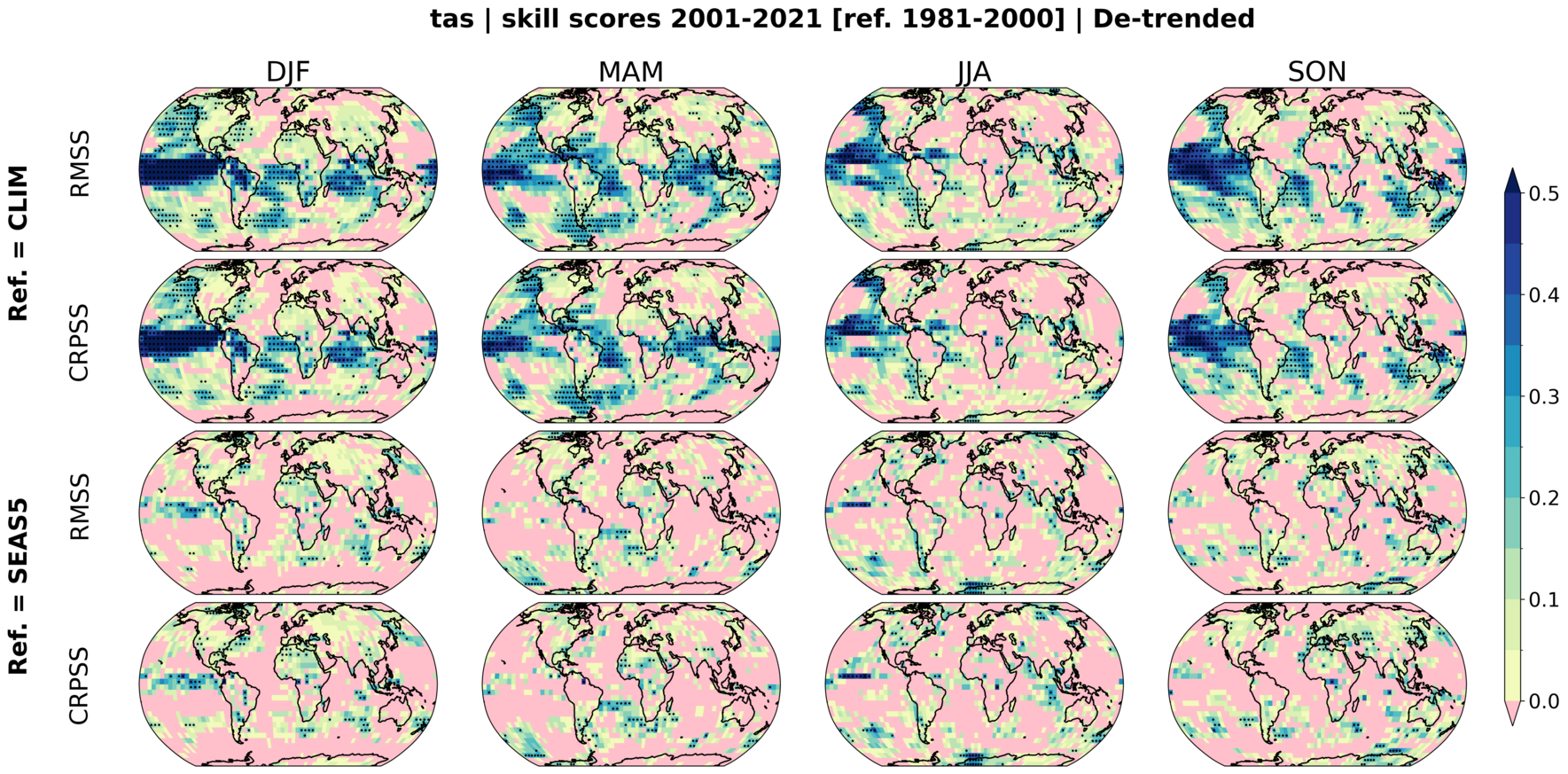}}
  \caption{\textbf{Global temperature skill scores.} Forecasts skill scores for near-surface air temperature (tas) predictions from 2001-2021, using 1981-2000 as the reference period of the climatological forecast and anomalies. The panels show four seasons (DJF, MAM, JJA, SON) across columns and two skill metrics: the root-mean-square error skill score (RMSS) and the continuous ranked probability skill score (CRPSS). Skill scores range from 0 (pink, indicating no skill) to 0.5 (dark blue, indicating high skill above the reference). Metrics are referenced (Ref.) against the climatological forecasts (CLIM) or against the ECMWF’s seasonal prediction system (SEAS5). Black dots indicate statistical significance of the skill score being positive at the 95\% confidence level  (more details can be found in the Methods section).}\label{f_glob_tas}
\end{figure}

\begin{figure}[!h]
 \centerline{\includegraphics[width=39pc]{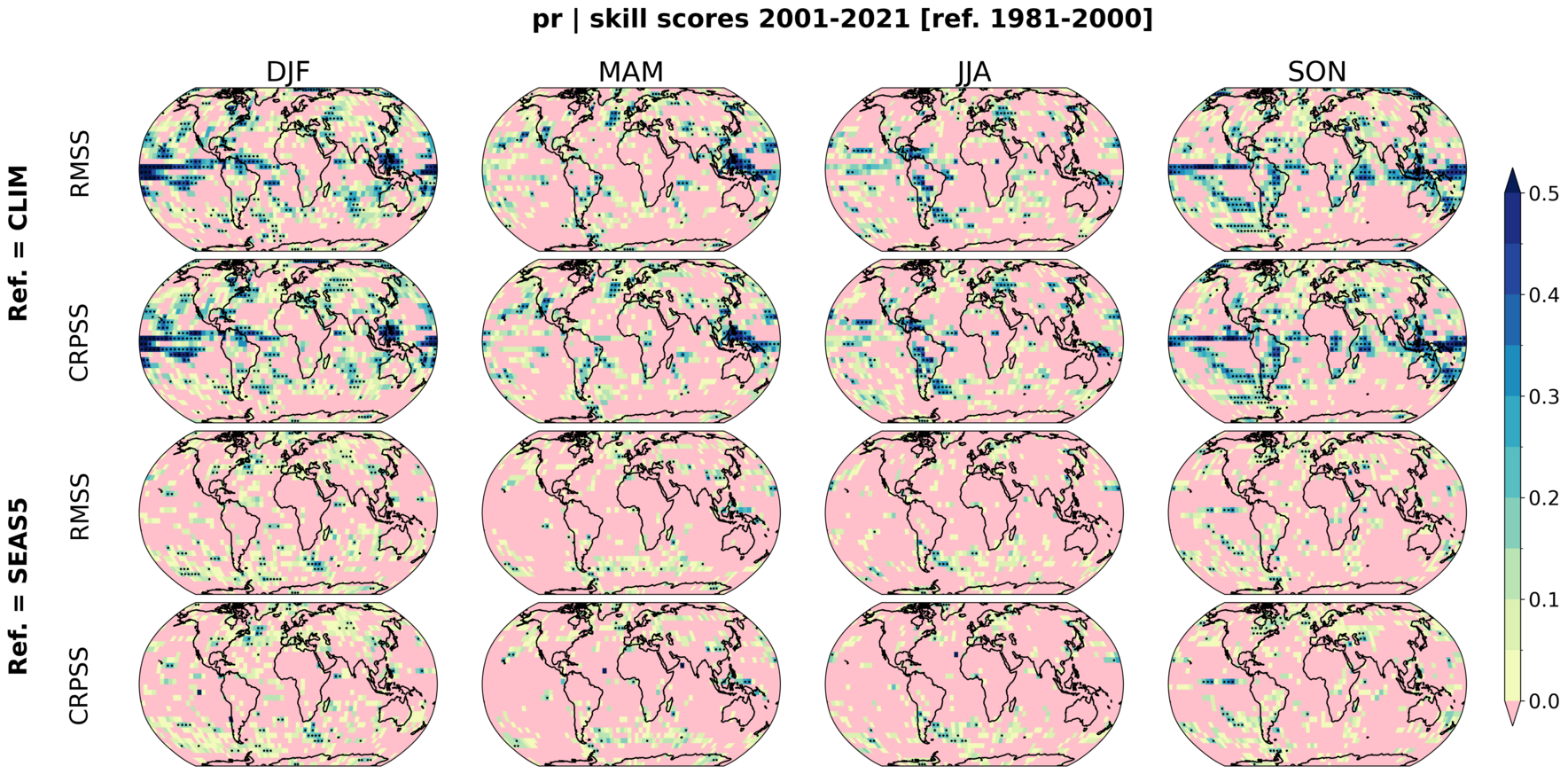}}
  \caption{\textbf{Global precipitation skill scores.} Same as Figure \ref{f_glob_tas} but for precipitation predictions.}\label{f_glob_pr}
\end{figure}

We compare our model with the climatological forecast (CLIM) and the ECMWF’s seasonal prediction system (SEAS5). The period covering 1981-2000 has been used as the reference period for the climatology and anomalies. Figure \ref{f_glob_tas} shows the forecast skill scores for near-surface air temperature (tas) predictions from 2001-2021. We present the results of four seasons (DJF, MAM, JJA, SON) across columns and using two skill metrics: the root-mean-square error skill score (RMSS) and the continuous ranked probability skill score (CRPSS). While the RMSS measures the deterministic performance of the ensemble median, the CRPSS offers a better view of the probabilistic performance of the ensemble distribution. Skill scores range from 0 (pink, indicating no skill) to 0.5 (dark blue, indicating high skill above the reference). 
Figure \ref{f_glob_pr} shows the results of precipitation predictions using the same two metrics. Black dots indicate whether the results are statistically significant at the 95\% confidence interval (more details in Methods).\\

Overall, the skill metrics relative to the climatological forecast exhibit similar patterns to those in Figure \ref{f_validation}, showing consistent performance across seasons. Temperature fields demonstrate higher and more spatially extensive skill compared to precipitation predictions. Similarly, for temperature, we observe a predominately more robust signal over the oceans compared to land regions and a clear ENSO and PDO signature. Skill over land areas is limited to regions such as the northern part of South America, Australia, central and northern Africa, and some parts of the US and Arctic regions. Precipitation forecasts also show a similar signal compared to the one observed in Figure \ref{f_validation}. We observe strong performance (Skill score above 0.2) compared to climatologies in regions such as northern South America, Australia, eastern and southern Africa, as well as parts of the U.S. and the Arctic. However, performance varies by season and region, with limited skill in Europe.
\\
We observe fewer regions showing improvements when comparing our model's performance against SEAS5 (third and fourth rows of Figures \ref{f_glob_tas} and \ref{f_glob_pr}). This is expected, as SEAS5 is the current state-of-the-art dynamical forecasting system, making it a more challenging benchmark to surpass compared to climatology. Generally, we observe that SEAS5 outperforms our approach in equatorial latitudes and most oceanic areas. Nevertheless, our model performs better in some land regions, which are particularly relevant for user applications. Regarding temperature forecasts, our approach shows improvements over SEAS5 in parts of the United States, the southern portion of South America, northwest Africa, Europe, and Eurasia (particularly in the SON season). In contrast, precipitation forecasts show little improvement against SEAS5. Yet, we observe season-dependent enhancements in some extra-tropical regions, i.e., Europe, Eurasia, parts of North America, and the most southern part of South America.\\

Should be noted that the added value from our predictions is only realized when there is an improvement over both SEAS5 and the climatological forecast. If this improvement is not achieved jointly, our data-driven approach may simply be converging to a simple climatological forecast. By comparing skill metrics for both references, we can identify areas where these two requirements are fulfilled. For temperature, such areas include parts of the United States, northwest Africa, the Arctic, and scattered locations in the rest of Africa and South America. For precipitation, we find few regions with improvements over SEAS5; mainly the extratropic regions, covering small parts of Europe, Eurasia, the US, and South America.

\subsection*{European regional use-case}\label{subsec:europe}

Once we tested the approach globally and verified that the model reproduces well-known patterns at the seasonal time scale, we verified the model with a Europe-centric context. As shown in Figure \ref{f1} (panel D), three configurations are tested, where models with the same inputs $X$ have three different targets $Y$: Global predictions at a spatial resolution of 5{\textdegree}, regional also at 5{\textdegree} and regional at 1{\textdegree}. The main objective is to assess whether targeting a more constrained region can yield improvements, as the model does not need to optimize its parameters for predicting the whole globe. Similarly, due to a smaller target domain, we can increase the spatial resolution of the target variable while maintaining the comparable computational cost needed for the original model (Global 5{\textdegree}). Thus, we assess whether we find benefits from increasing the spatial resolution of the predictions. Additionally, we test whether the patterns are consistent across the different configurations.\\

\begin{figure}[!h]
 \centerline{\includegraphics[width=39pc]{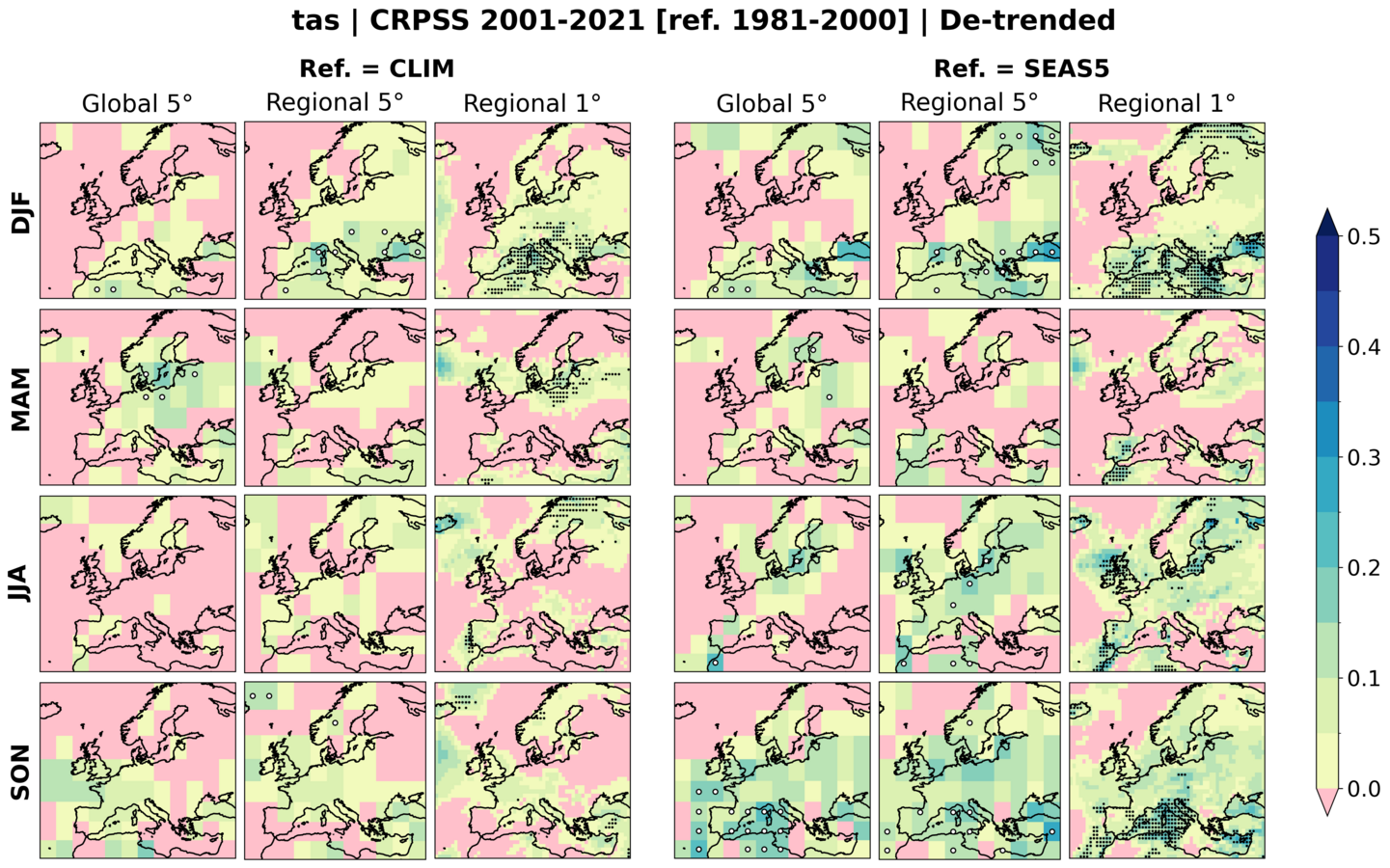}}
  \caption{\textbf{Regional temperature CRPSS.} Temperature forecasts skill scores across seasons (DJF, MAM, JJA and SON) of the three model configurations described in Figure \ref{f1}, panel D: Global 5{\textdegree}, Regional 5{\textdegree} and Regional 1{\textdegree}. Metrics are referenced (Ref.) against the climatological forecasts (CLIM) or against the ECMWF’s seasonal prediction system (SEAS5). Dots indicate statistical significance that the skill score is positive at the 95\% confidence level.}\label{f_eu_tas}
\end{figure}

The CRPSS for temperature predictions over Europe shows varying degrees of skill compared to both climatology and SEAS5. As a first assessment, we can observe across the different models and seasons higher skill values against the climatology than SEAS5, indicating the poor performance of SEAS5 in Europe. We also observe consistent patterns across model configurations, with the Regional 1{\textdegree} offering the most detailed spatial representation of skill, highlighting statistically significant improvements in specific regions while still aligning with patterns observed at coarser resolutions. Across the different configurations we observe how the model manages to improve the predictions over SEAS5 and the climatology for large parts of the Central and Western Mediterranean (DJF and SON), eastern Mediterranean (JJA), and also some parts of Northern and Central Europe (MMA). As mentioned in the previous section, improvements over SEAS5 and the climatology are needed to add value to user-centered applications.\\

%Figures \ref{f_eu_tas} and \ref{f_eu_pr} present skill scores identical to those in Figures \ref{f_glob_tas} and \ref{f_glob_pr} but for the new setup. For temperature, significant skill against climatology is observed across large parts of Europe during DJF and JJA, while skill against climatological forecasts is limited to a few areas in MAM (Balkans) and SON (Northern Europe). Compared to SEAS5, more extensive positive skill signals are evident, particularly in JJA and SON. When jointly evaluated against both references, our approach shows added value in Northern Europe (JJA and SON), parts of Italy (SON and MAM), and scattered regions for the DJF season. Regarding precipitation (figure 7), we observe widespread positive skill patterns against the climatological forecast, with notably higher values in JJA, especially in areas such as the Iberian Peninsula, north of the Black Sea, and along the east and west coasts of the North Sea. Compared to SEAS5, our method demonstrates spatially extensive patterns of positive skill across regions, aligning with the positive results obtained against the climatological forecast. Consequently, the added value of our approach is more pronounced and extends to most of Europe for precipitation forecasts, in contrast to the more limited improvements seen in temperature predictions.\\

\begin{figure}[!h]
 \centerline{\includegraphics[width=39pc]{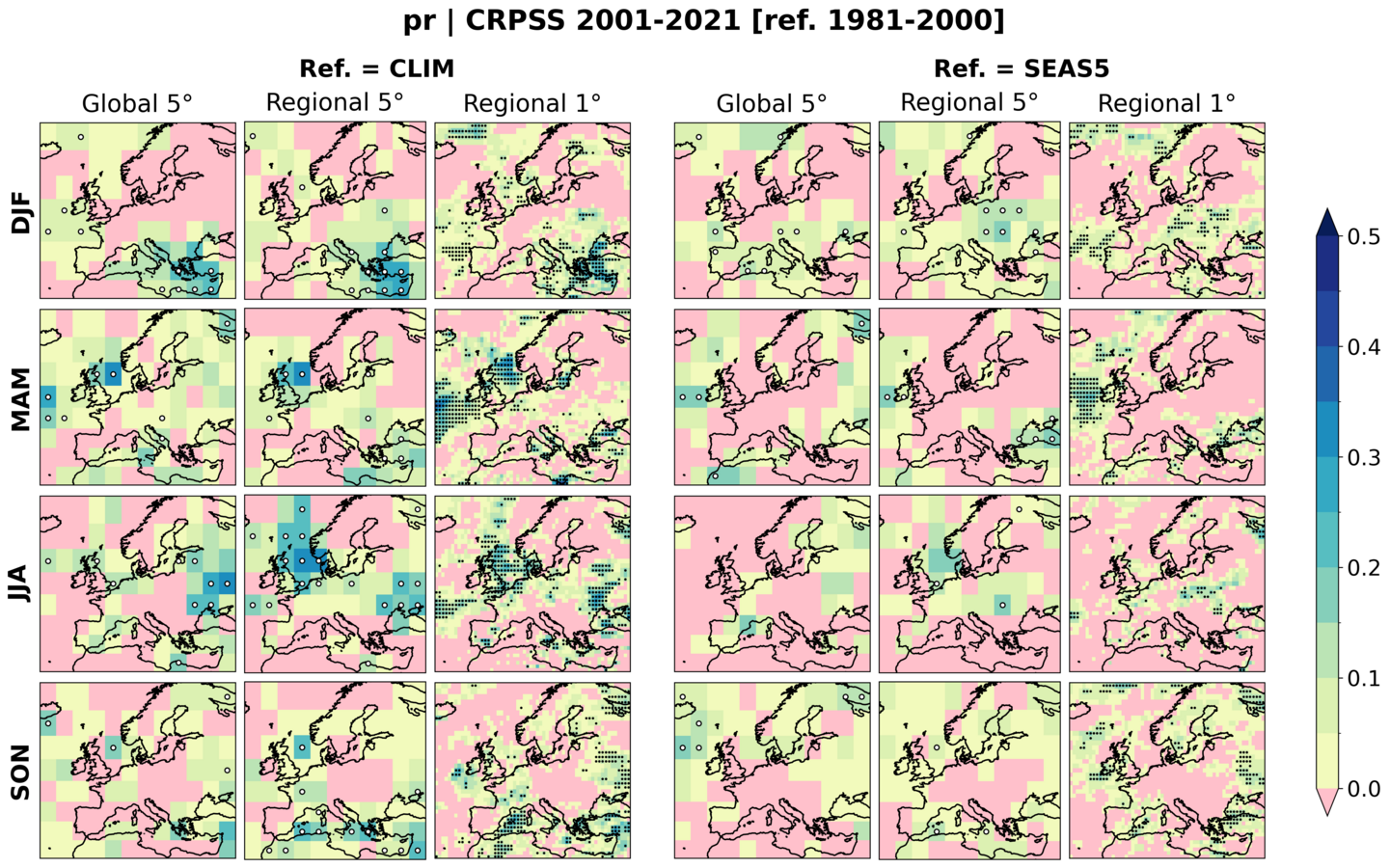}}
  \caption{\textbf{Regional precipitation CRPSS.} Same as Figure \ref{f_eu_tas} but for precipitation predictions. 
}\label{f_eu_pr}
\end{figure}

Contrary to what has been observed for temperature forecasts, we observed in general terms (across seasons and models) higher precipitation skills against climatology than those obtained against SEAS5. SEAS5 seems to offer a more competitive performance compared to our approach and compared to temperature forecasts. Still, our approach improves SEAS5 performance even further over a big set of regions, and across seasons. We observe relevant improvements both against the climatology and SEAS5 over parts of central to eastern Europe (DJF and JJA), parts of the British Isles (MAM, SON and JJA), and northern Europe (SON). Again, we observe how the general patterns are maintained against the different configurations, with the 1{\textdegree} configuration highlighting a more detailed but coherent spatial representation of skill and significance.\\

The robustness of the results across various spatial configurations (Global 5°, Regional 5°, and Regional 1°) represents an interesting finding. The skill patterns remain consistent regardless of the domain's resolution or setup. In addition, we observe that in most cases, the skill is either retained or improved by going regional or increasing the spatial resolution. Thus, whether which one is the most suitable solution might depend then on user and computational requirements. Thus, the presented approach improves predictions across many regions in Europe and showcases how targeting regional domains can be a cost-effective solution for creating predictions at a higher spatial resolution while retaining a global context. \\

%Figures \ref{f_eu_tas} and \ref{f_eu_pr} compare the CRPSS for temperature and precipitation between the model described in the previous section and a new model optimized for Europe (also illustrated in Figures 6 and 7). It is important to note that both models utilize identical inputs at a 5°x5° resolution. Despite being independently trained from scratch, the two models exhibit remarkably similar skill patterns across seasons, metrics, and variables, with few exceptions. This similarity suggests that the method converges to a similar degree of predictability even when targeting a more constrained region. While this convergence lends robustness to the approach, it also indicates that, in this case, higher resolution on the target or a more parametrized model does not necessarily enhance performance. These findings expose whether a predictability limit has been reached with the current configuration of input variables and climate model output used for training.

\section*{Discussion}\label{sec:discusion}

This study evaluates the performance and robustness of a probabilistic deep learning method in predicting global seasonal anomalies at a one-month lead, aligning with the setup of most climate services providers. The model has been trained on climate model output from CMIP6 and tested using ERA5 reanalysis data, explicitly accounting for climate change trends. It integrates variational inference with vision transformers, representing the first application of transformers for seasonal prediction to our knowledge. We also apply the same methodology to a reduced target domain focus in Europe.\\

Our results reveal distinct patterns of prediction skill across different variables and regions. Temperature predictions demonstrate skill beyond the climate-change trend and outperform SEAS5 in various inland areas, though SEAS5 maintains superior performance in most of the equatorial band. Precipitation forecasts show more limited skill, with fewer regions outperforming climatology and fewer surpassing SEAS5. Notably, the generative capabilities enabled by conditional variational autoencoders allow efficient generation of ensemble members at reduced computational cost, effectively filtering the predictable signal, as previously observed by \citep{panImprovingSeasonalForecast2022}.\\

The model's predictive skill beyond trend-based forecasts validates our methodological approach and the utility of climate model output in capturing interannual seasonal variability. The superior performance in inland areas compared to SEAS5 suggests that our model might better represent links between inputs and targets, but it is unclear whether this is due to land surface-atmosphere interactions or teleconnection-related processes. Similarly, the consistency of skill patterns across different target configurations reinforces the robustness of our methodology, indicating that these patterns represent genuine features in the climate system rather than artefacts of the machine learning model.\\

Nevertheless, important limitations emerge from our analysis. SEAS5's superior performance in equatorial regions likely stems from its more refined representation of tropical ocean and atmosphere dynamics and their coupling \citep{Johnson2019}. Our approach's limitations in these regions may arise from the input variables used to define the climate state, the biases in the climate model outputs used to fit the model or inherent constraints of the deep learning methodology itself \citep{yacobyFailureModesVariational2022}. Similar conclusions can be extrapolated to precipitation predictions, where skill remains modest. Additionally, our use of monthly and seasonal means for both input and target data differs significantly from the daily to sub-daily initialization processes in dynamical systems \citep{Doblas-Reyes2013a}, potentially limiting our ability to capture rapid atmospheric changes. Addressing the importance of information at different timescales and how this is propagated in deep learning and dynamical models is a line of research worth further investigation. Finally, while our validation metrics partially address forecast reliability \citep{ReliabilitySeasonalClimate}, this aspect has not been explicitly addressed in this work and warrants further analysis. Generative models offer an explicit framework for predicting the conditional likelihood of future climate states. However, whether this approach provides more reliable uncertainty quantification than traditional ensembles based on perturbed initial conditions remains an open question. \\

This work's implications are substantial for operational forecasting and methodological advancement. The model's computational efficiency in ensemble generation could enable more frequent forecast updates and larger ensemble sizes in operational settings. The demonstrated skill in European forecasts suggests promising potential for regional-specific implementations, potentially enhancing local climate services. Furthermore, the successful application of vision transformers represents an important methodological advance, demonstrating their capability to capture complex spatial dependencies in climate data. This study advances our initial objective of further developing probabilistic deep learning methods for seasonal prediction \citep{panImprovingSeasonalForecast2022}, demonstrating that AI-based approaches can achieve comparable skill to operational dynamical systems in many regions. While challenges remain in matching the performance of numerical prediction systems in specific regions, our results establish a promising foundation for the future development of hybrid forecasting systems that leverage the strengths of both data-driven and dynamical approaches.

\section*{Methods}\label{sec:methods}

\subsection*{\textbf{Problem formulation}}
The objective is to predict the climate state $\mathbf{Y} \in \mathbb{R}^{c_y \times n_{lat} \times n_{lon}}$ of a future season based on current and past states $\mathbf{X}^i \in \mathbb{R}^{c_x \times n_{lat} \times n_{lon}}$ from the $i$ preceding months. To deal with the stochastic nature of the atmosphere beyond 12 days \citep{lorenzDeterministicNonperiodicFlow1963} we intend to forecast not a deterministic value but the conditional probability distribution $p(\mathbf{Y}^{t+1} | \mathbf{X}^{t},\mathbf{X}^{t-1},...,\mathbf{X}^{t-T})$ of the target season $\mathbf{Y}^{t+1}$ on the current and previous conditions $\mathbf{X}$. \\

The representation of the target season $\mathbf{Y}$, is comprised by $c_y$ variables of 3-month seasonal averages on a 5{{\textdegree}}{$\times$}5{{\textdegree}} latitude-longitude grid. As stated in \citep{goddardVerificationFrameworkInterannualdecadal2013}, this grid-scale is a good compromise between capturing the large-scale climate signal and smoothing out noise while saving computational resources. The representation of the initial states $\mathbf{X}$, is comprised by $c_x$ variables of monthly averages on a 5{{\textdegree}}{$\times$}5{{\textdegree}} latitude-longitude grid. The same methodology can be tested under different representations, combining different grids and temporal resolutions (both at source and output) due to its inherent flexibility. Thus for the regional use case, we increase the spatial resolution of the target $\mathbf{Y}$ to 1{{\textdegree}} (see Section \ref{sec:results}\ref{subsec:europe}).

\subsection*{\textbf{Variational Inference}}

To obtain the conditional probability distribution $p(\mathbf{Y}|\mathbf{X})$ on a target climate state $\mathbf{Y}$ given a state $\mathbf{X}$ of current or past conditions, state-of-the-art climate prediction systems run multiple dynamical simulations, each with slightly perturbed initial conditions, obtaining an ensemble of plausible outcomes from which probabilities can be inferred. Analogously, our objective is to learn a statistical model $p_\theta(y|x,z)$ from which multiple predictions can be inferred statistically from a set of initial conditions and an n-dimensional latent variable $z$  that adds the stochastic component to the statistical model.\\

Learning the conditional probability distribution  $p_\theta(y|x)$ from data is not a straightforward problem \citep{murphyProbabilisticMachineLearning2023}. Ideally, we would like to minimise the difference between our learned distribution $p_{\theta}(y|x)$ and the observed data distribution $q_D(y|x)$. This objective can be achieved empirically by maximising the sum over the log-likelihoods of our data points in the learned distribution \citep{wilksStatisticalMethodsAtmospheric2019}. Yet, this is computationally intractable as it requires integration over $z$ for each data point: 
\begin{equation}
p_\theta(y|x) = \int_z p_\theta(y|z,x)p_\theta(z|x)dz 
\end{equation}

Amortised variational inference \citep{kingmaAutoEncodingVariationalBayes2013} offers an alternative by narrowing the integration space of z to values that are likely to generate y. This likelihood is described by $p(z|y_i,x_i)$ and is approximated by an amortised inference distribution $q_\phi(z|y,x)$ that is also learned. To jointly optimise the parameters $\phi$ and $\theta$  a lower-bound of the log-likelihood or evidence lower-bound (ELBO) is defined:
\begin{equation}
     \mathcal{L}(\theta, \phi) = -\mathbb{E}_{q_\phi(z|x,y)}[\log p_\theta(y|z,x)] + D_{KL}(q_\phi(z|y,x) || p_\theta(z|x))
\end{equation}

where $\mathbb{E}_{q_\phi(z|x,y)}[\log p_\theta(y|z,x)]$ is the expected log-likelihood of $y$ given $z$ and $x$, and $D_{KL}(q_\phi(z|y,x) || p_\theta(z|x))$ is the KL divergence between the approximate posterior $q_\phi(z|y,x)$ and the prior $p_\theta(z|x)$. \\

Thus, our final objective is to jointly train two neural networks: $q_\phi(z|y,x)$ representing the learned approximate posterior, and $p_\theta(y|z,x)$ being the learned generative model. $q_\phi(z|y,x)$ will be represented by an encoder applied to the target state $\mathbf{Y}$, and only used during the training phase. While $p_\theta(y|z,x)$ will be conformed by an encoder on the initial state $\mathbf{X}$ and the decoder generating new predictions $\mathbf{Y_\theta}$ combining information from the learned latent space and the compact representation of the initial state $\mathbf{X}$. Minimising this ELBO allows joint optimisation of $\theta$ and $\phi$, effectively approximating the intractable $p_\theta(y|x)$.
%- not sure if the equation of the ELBO is the best
%- also there must be some mentioning to stochastic version using batches of samples?

\subsection*{\textbf{Architecture}}

The model architecture design is essential for extracting meaningful features that improve seasonal predictability. The architecture needs to capture both temporal and spatial long-range interactions influenced by global teleconnections, as well as local interactions that stem from land-atmosphere processes and persistence. However, due to the limited size of the available training dataset, keeping the model complexity in check is essential to avoid overfitting. Choosing the architectural design implies finding a sustainable balance between these factors.\\

Vision Transformers (ViTs) are a well-suited option for this task \citep{vaswaniAttentionAllYou2023, dosovitskiyImageWorth16x162021}. ViTs employ a general-purpose inductive bias that allows them to model distant and local connections without needing the deep hierarchy and pooling operations typical of Convolutional Neural Networks (CNNs). Thus, they decouple the interaction range from the network depth, and this is particularly helpful when modelling the different types of interactions that occur at seasonal time scales. In addition, transformers are very suitable for incorporating data with different formats (i.e. time series with 2D or 3D spatial grids). They can also make inferences even under the erratic presence of missing values. Still, due to their unconstrained non-locality, ViTs are known to need large datasets in order to train correctly. These reasons partly explain the multiple applications of ViTs found in weather prediction \citep{bodnarAuroraFoundationModel2024,nguyenScalingTransformerNeural2023,nguyenClimaXFoundationModel2023}, contrasting the few to no applications for seasonal prediction.\\

As illustrated in Figure~\ref{f1} panel B, our model architecture combines the variational inference framework of a conditional Variational Autoencoder (cVAE) with ViT encoders for feature extraction. The $q_\phi(z|y,x)$ approximate posterior is represented by a ViT encoder applied to the target climate state $\mathbf{Y}$. This encoder generates a compact latent representation $\mathbf{z_y}$, which is then passed through a Multi-Layer Perceptron (MLP) to produce the variational parameters $\mu$ and $\sigma$. This network component (depicted in orange) is only used during training. The $p_\theta(y|z,x)$ generative model is formed by an additional ViT encoder applied to the initial climate state $\mathbf{X}$, producing a reduced representation $\mathbf{z_x}$. This reduced representation  $\mathbf{z_x}$ latent is then combined with a sampled latent $\mathbf{z}$ from the prior and passed through a Convolutional Neural Network (CNN) decoder to generate new climate predictions $\mathbf{Y_\theta}$. During inference, the latent distribution $\mathbf{z}$ is sampled multiple times to obtain different predictions that form an ensemble. \\

This architecture allows the model to extract meaningful features from the input data while maintaining a constrained overall size. By jointly optimizing the encoder ($q_\phi$) and decoder ($p_\theta$) networks using minimizing loss objective, the model learns to generate diverse, physically consistent ensemble predictions while capturing the underlying uncertainty in the data. 

%\*section{\textbf{Experimental Setup}} \label{sec:exp_setup}

\subsection*{\textbf{Datasets}} \label{datasets}

We use four different climate models (see Table~\ref{tab:cmip6-dataset}) from the Coupled Model Intercomparison Project 6 \citep[CMIP6]{eyringOverviewCoupledModel2016} to obtain a sufficiently large training set. The historical and SPP2-4.5 scenarios are concatenated for each realisation into a continuous time series spanning 1880 to 2080. These specific models were chosen as they meet the criteria for the number of realisations and output variables. All the models' output was obtained from the Earth System Grid Federation (ESGF) and gathered and pre-processed to joint spatial resolution and units using ESMValtool \citep{beobide-arsuagaSpringRegionalSea2023}.\\

\begin{table}[htbp]
\centering
\begin{tabular}{|l|l|l|l|}
\hline
Split & Source & Time Period & Models \\
\hline
Training & CMIP6 (Hist. + SSP245) & 1880 - 2080 & \begin{tabular}[c]{@{}l@{}}CanESM5\_r(6:25)i1p1f1, CanESM5\_r(6:25)i1p2f1,\\ MIROC-ES2L\_r(6:25)i1p1f2, MIROC6\_r(6:25)i1p1f1 \&\\ MPI-ESM1-2-LR\_r(6:25)i1p1f1\end{tabular} \\
\hline
Validation & CMIP6 (Hist. + SSP245) & 1880 - 2080 & \begin{tabular}[c]{@{}l@{}}CanESM5\_r(1:5)i1p1f1, CanESM5\_r(1:5)i1p2f1,\\ MIROC-ES2L\_r(1:5)i1p1f2, MIROC6\_r(1:5)i1p1f1 \&\\ MPI-ESM1-2-LR\_r(1:5)i1p1f1\end{tabular} \\
\hline
Test & ERA5 & 1950 - 2021 & \\
\hline
\end{tabular}
\caption{Datasets description information}
\label{tab:cmip6-dataset}
\end{table}

For the evaluation of the data-driven models, we use the ERA5 \citep{Hersbach2020} reanalysis covering 1950 to 2021. ERA5 is produced using 4D-Var data assimilation combined with the ECMWF Integrated Forecast System (IFS) CY41R2. Again, ERA5 data was preprocessed to a common spatial grid and units using ESMValtool.\\

%(Observational data?)\\

We also use the ECMWF's seasonal climate prediction SEAS5 \citep{Johnson2019} as a dynamical benchmark against the data-driven models. SEAS5 is based on the Integrated Forecast System (IFS) atmospheric component coupled to the Nucleus for European Modelling of the Ocean (NEMO) ocean model and the dynamic Louvain‐la‐Neuve Sea Ice Model (LIM2). SEAS5 operational seasonal forecasts are initialised on the first day of each month, and 51 ensembles are initialised covering up to seven months in the future. Additionally, a set of hindcasts (1981 to 2017) are also produced with the same configuration but with a reduced ensemble (25 realisations). As a benchmark, we concatenate both hindcast and forecasts into a continuous set of forecasts covering the period 1981 to 2021 with an ensemble of 25 realisations. The SEAS5 data was obtained from the Copernicus Climate Data Store (CDS) API.

\subsection*{\textbf{Data preprocessing}}

We perform an initial homogenization of all the climate model output and reanalysis data to a common spatio-temporal resolution and units. For both inputs and outputs, monthly means at 5{\textdegree} or 1{\textdegree} horizontal resolution are used (depending on the prediction task).\\

The second stage involves the standardisation of the data, including de-trending, anomaly computation, and normalisation. Data standardisation is a critical aspect of data-driven climate predictions. On one hand, the standardisation of inputs and outputs can drastically change the prediction task assigned to the model (i.e. predicting seasonal averages over the trend vs predicting the forcing influence at the seasonal time scale). Similarly, it can affect the models' performance, as the standardisation can remove or add information from the climate fields used. Finally, improper standardisation of the outputs, references, and benchmarks can lead to misleading claims of performance during the validation \citep{risbeyStandardAssessmentsClimate2021a}, especially under strong trends \citep{goddardVerificationFrameworkInterannualdecadal2013, tippettTrendsSkillSources2024}. At the same time, data standardisation helps in the speed and stability during the training of data-driven approaches.\\

De-trending for CMIP6 data is performed removing the forced component (ensemble mean) of each model independently. For the ERA5 reanalysis, locally estimated scatterplot smoothing (LOESS)  \citep{mahlsteinEstimatingDailyClimatologies2015}, with a fixed time window of 30 years and one degree of freedom, is applied to obtain a non-linear trend later removed from the data. To avoid overestimates of forecast skill due to the use of information not available at forecast time \citep{risbeyStandardAssessmentsClimate2021a}, we fit LOESS using only values prior to the forecasting time (retroactively). As shown in Figure \ref{f1}, de-trending is applied to the target Y during training, as well as during the validation of the forecasts.\\

Standardised monthly anomalies are computed by subtracting the mean and normalising by the standard deviation of the 1981 to 2000 period. As an exemption, precipitation values are fitted to a gamma distribution instead. All these steps are applied point by point to each climate model and reanalysis independently, helping to remove significant biases present in both climate models and the reanalysis output. However, none of the data from the testing period 2001-2021 is included in this process.

\subsection*{\textbf{Assessing forecasts quality}}

In this work, we use a set of verification metrics to quantify the quality of the predictions developed and we compare the results against the ECMWF state-of-the-art seasonal prediction system SEAS5. As exposed in \citep{goddardVerificationFrameworkInterannualdecadal2013}, we identify two main objectives. First, to assess whether our proposed model produces more accurate predictions compared to a reference forecast, in our case a state-of-the-art dynamical forecasting system. Second, to assess whether the ensemble spread of our method provides a good estimation of uncertainty on average.\\

The first objective can be fulfilled by employing deterministic metrics. As a first individual assessment of the different forecasts, we employ the Spearman correlation (Equation 1) between the anomalies of the ensemble median of our predictions and the ground truth, also known as the Anomaly Correlation Coefficient (ACC). It helps quantify the monotonic relationship between these two. The Spearman correlation is preferred over the Pearson correlation due to its non-parametric nature and insensitivity to outliers.

\begin{equation}
\rho = 1 - \frac{6 \sum_{i=1}^n (r_{x_i} - r_{y_i})^2}{n(n^2 - 1)}
\end{equation}

where $r_{x}$ is the ranks of the predictions ensemble median, $r_{y}$ the ranks of the ground truth, and $n$ the number of samples.\\

In addition, the root mean square error (RMSE) is used to add information of the potential mean and conditional biases in our predictions (Equation 2). The RMSE can be expressed as a function of the Spearman correlation and the mean and conditional biases \citep{murphyalanSkillScoresBased1988}, providing a complete deterministic overview of our predictions.

\begin{equation}
\text{RMSE} = \sqrt{\frac{1}{n} \sum_{i=1}^n (\hat{x}_i^2 - y_i^2)}
\end{equation}

where $\hat{x}$ is the ensemble median of our predictions, $y$ are the observations, and $n$ the number of samples.\\

Our second objective is better fulfilled using probabilistic metrics, which test whether the spread in our prediction is adequate to quantitatively represent the range of possibilities for individual predictions over time. We base our probabilistic validation on the Continuous Ranked Probability (CRPS), a measure of squared error in the probability space:

\begin{equation}
\text{CRPS}(P, y) = \int_{-\infty}^{\infty} [F(x) - H_y(x)]^2 dx 
\end{equation}

where $F$ is  the proposed cumulative distribution function (CDF) obtained from the forecast ensemble $x$, and $H$ is the Heaviside step function centered at the actual observed value $y$.\\

To facilitate the interpretability and comparison of the results, both CRPS and RMSE are expressed as skill scores referenced against a climatological forecast (clim) or SEAS5:
\begin{displaymath}
CRPSS = 1 - \frac{CRPS}{CPRS_{ref}} \quad
RMSS = 1 - \frac{RMSE}{RMSE_{ref}}
\end{displaymath}

Uncertainty in the validation metrics is evaluated using a non-parametric bootstrapping approach. The forecasts and reanalysis observations are reshuffled in this method to compute 150 core values. For the ACC computation, the values obtained are compared to the 95th significance level against a similar distribution generated using a random time series instead of the forecast. For the skill score metrics, we reshuffled the forecast, reference forecast, and reanalysis time series to compute a distribution of skill scores. Then, we assess whether the score value is significantly greater than zero at the 95th significance level.

\subsection*{\textbf{Architecture \& training configuration}}

The model processes five key climate variables: 2-meter air temperature (tas), precipitation (pr), sea surface temperature (tos), and geopotential height at 500hPa and 300hPa levels (zg500, zg300). For the input state X, these variables represent conditions during the preceding 6 months, while the target state Y is comprised of the same variables seasonal averaged comprised by lead months 1 to 3. Topography, land-ocean, and encoded latitude and longitude coordinates are also concatenated to the input state X. During inference, 150 ensembles are pooled by sampling from the latent space and conditioned on the inputs.\\
 
Our conditional Variational Autoencoder (cVAE) implements dual Vision Transformer (ViT) encoders to process input and target climate states. Each encoder pathway consists of 8 transformer layers with an embedding dimension of 128 and single-head attention, operating on patch sizes of 1 to capture fine-grained spatial features. The latent space has a dimension of 128, enabling a compact representation of climate patterns. The decoder employs four residual blocks with 32 filters each, using convolutional layers to reconstruct the predicted climate fields.\\

We train an independent model for each season. Each one underwent training for 100 epochs with a batch size of 32, using an initial learning rate of $5e^{-5}$ and weight decay of 0.01 for regularization. The model is optimized through an information maximization loss function for variational autoencoders, or InfoVAE objective \citep{zhaoInfoVAEInformationMaximizing2018}, that constitutes a generalization of the ELBO objective. The weighting parameter (lambda) was set to 1 and the confidence parameter (alpha) to 0.95. This configuration was selected according to the results obtained in the validation set when sampling pseudo-randomly different hyper-parameter configurations (not shown).
\\
The basic configuration and hyper-parameters of the model are consistent across the experiments shown in panel D of Figure \ref{f1}, with one exception: in the Regional 5{\textdegree} model, the number of blocks in the decoder is reduced from 4 to 2 due to the limited number of points. Additional details regarding the architecture and hyper-parameters can be found in Supplementary \ref{Appx_b}.

%%%%%%%%%%%%%%%%%%%%%%%%%%%%%%%%%%%%%%%%%%%%%%%%%%%%%%%%%%%%%%%%%%%%%
% DATA AVAILABILITY STATEMENT
%%%%%%%%%%%%%%%%%%%%%%%%%%%%%%%%%%%%%%%%%%%%%%%%%%%%%%%%%%%%%%%%%%%%%
% 
%
%\datastatement
 
%All the data used are publicly available or restricted to the signed-up users. SEAS5 and ERA5 data were downloaded from the official website of Copernicus Climate Data (CDS) at https://cds.climate.copernicus.eu/. CMIP6 datasets were downloaded from the Earth System Grid Federation (ESGF). Pre-processed Zarr datasets used to train the models will be made publicly available upon acceptance of this manuscript.

%The underlying code for this study and training/validation Zarr datasets are not publicly available but may be made available upon request from the corresponding author.

%The code used to generate the results presented in the paper as well as pre-trained model weights and analysis code for generating plots will be made publicly available (with persistent identifier) upon acceptance of this manuscript.

%Authors contribution
%Ll.P., A.M. and A.P. conceived the research idea. Ll.P. and A.P. implemented the deep learning code. Ll.P. Implemented the validation pipeline. D.C. contributed to the development of the code. Ll.P. drafted the manuscript with input from all co-authors. All authors discussed the results and revised the manuscript. M.D., J.P., L.R. and A.S. supervised the project.

%%%%%%%%%%%%%%%%%%%%%%%%%%%%%%%%%%%%%%%%%%%%%%%%%%%%%%%%%%%%%%%%%%%%%
% REFERENCES
%%%%%%%%%%%%%%%%%%%%%%%%%%%%%%%%%%%%%%%%%%%%%%%%%%%%%%%%%%%%%%%%%%%%%
\clearpage
\bibliographystyle{plainnat}
\bibliography{references}

%%%%%%%%%%%%%%%%%%%%%%%%%%%%%%%%%%%%%%%%%%%%%%%%%%%%%%%%%%%%%%%%%%%%%
% ACKNOWLEDGMENTS
%%%%%%%%%%%%%%%%%%%%%%%%%%%%%%%%%%%%%%%%%%%%%%%%%%%%%%%%%%%%%%%%%%%%%
%\acknowledgments
%This work was supported by the AI4Drought project (ESA AI4SCIENCE; contract number 4000137110/22/I-EF). AD holds a fellowship within the "Generación D" initiative, Red.es, Ministerio para la Transformación Digital y de la Función Pública, for talent attraction (C005/24-ED CV1), funded by the European Union NextGenerationEU funds, through PRTR. MGD and SM are grateful for support from the Horizon Europe project EXPECT (Grant 101137656). The authors thank Pierre-Antoine Bretonnier and Margarida Samsó for downloading and pre-processing the datasets used in this project.

%%%%%%%%%%%%%%%%%%%%%%%%%%%%%%%%%%%%%%%%%%%%%%%%%%%%%%%%%%%%%%%%%%%%%
% APPENDIXES
%%%%%%%%%%%%%%%%%%%%%%%%%%%%%%%%%%%%%%%%%%%%%%%%%%%%%%%%%%%%%%%%%%%%%

%% If only one appendix, use

%\appendix

%% If more than one appendix, use \appendix[<letter>], e.g.,

%\appendix[A] 

%% Appendix title is necessary! For appendix title:

%\appendixtitle{Title of Appendix}

%%% Appendix section numbering (note, skip \section and begin with \subsection)
%
% \subsection{First primary heading}

% \subsubsection{First secondary heading}

% \paragraph{First tertiary heading}

\appendix[A] 

\appendixtitle{Supplementary figures}

\subsection*{Results}

\begin{figure}[h!]
 \centerline{\includegraphics[width=39pc]{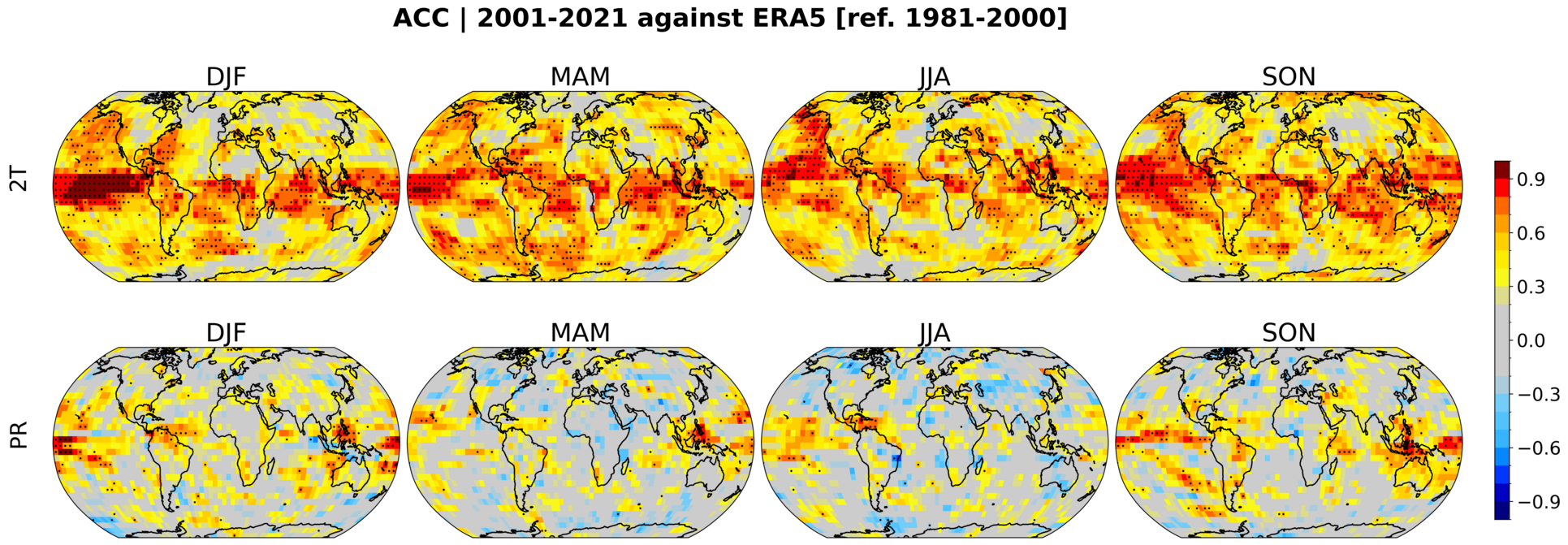}}
  \caption{
  Same as Figure \ref{f_validation} but without de-trending.
}\label{f_A1}
\end{figure}

\begin{figure}[h!]
 \centerline{\includegraphics[width=39pc]{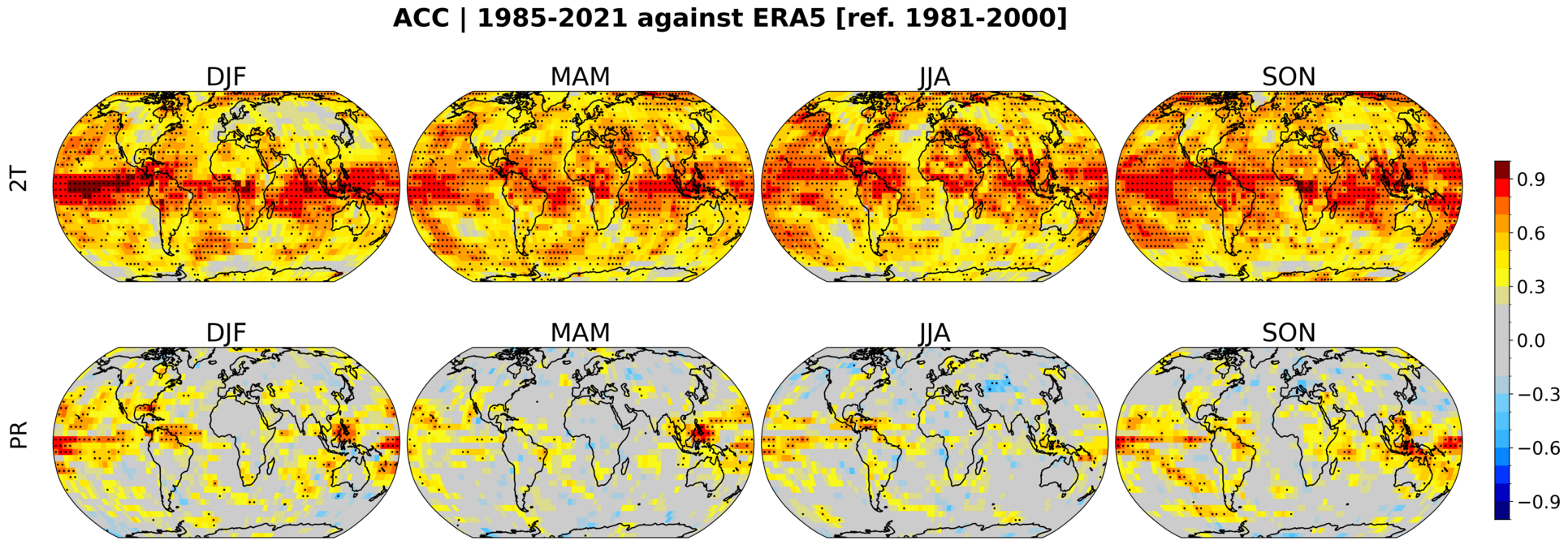}}
  \caption{
   Same as Figure \ref{f_validation} but without de-trending and covering the period 1985-2021.
}\label{f_glob_acc_det_extended}
\end{figure}

\begin{figure}[h!]
 \centerline{\includegraphics[width=39pc]{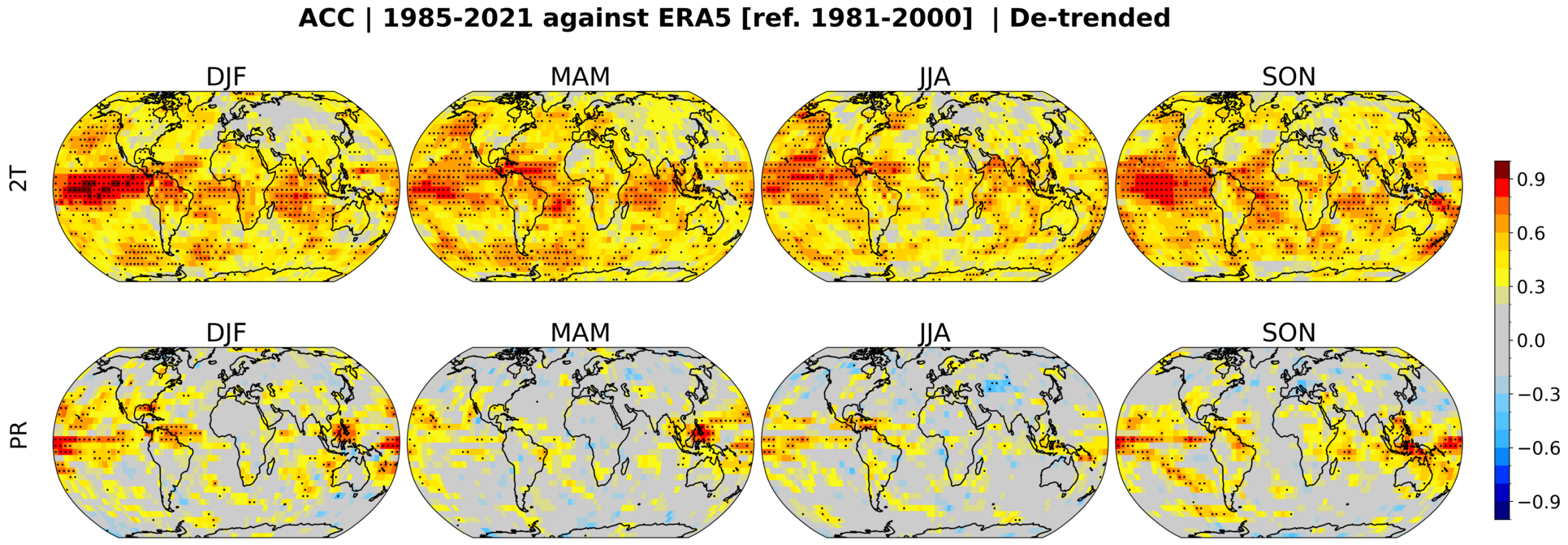}}
  \caption{
     Same as Figure \ref{f_validation} but covering the period 1985-2021.
}\label{f_glob_acc_det_extended}
\end{figure}

\begin{figure}[!h]
\centerline{\includegraphics[width=39pc]{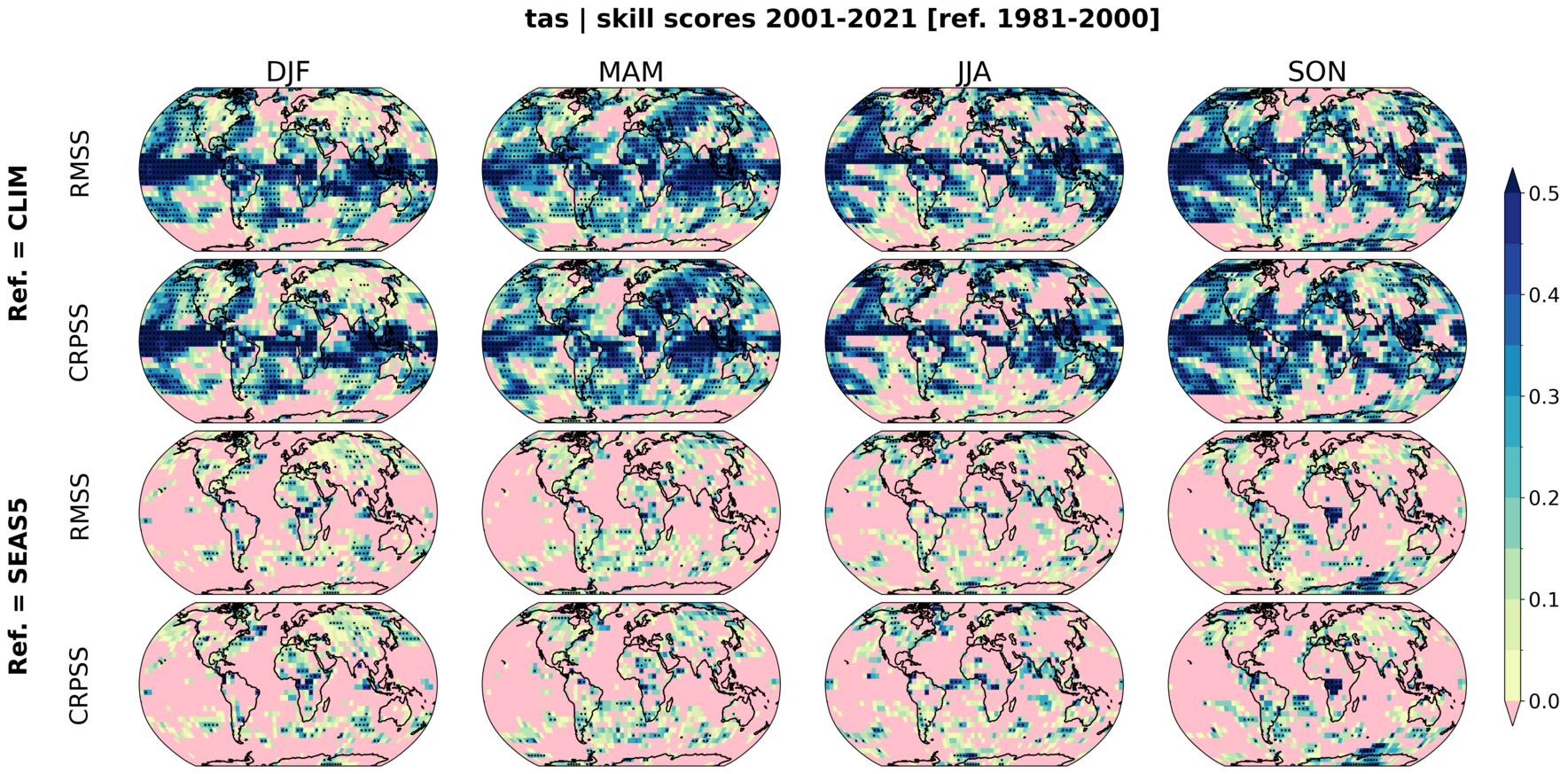}}
  \caption{Same as Figure \ref{f_glob_tas} but without de-trending.
}\label{f_glob_skill_tas_nodet}
\end{figure}

\begin{figure}[!h] 
\centerline{\includegraphics[width=39pc]{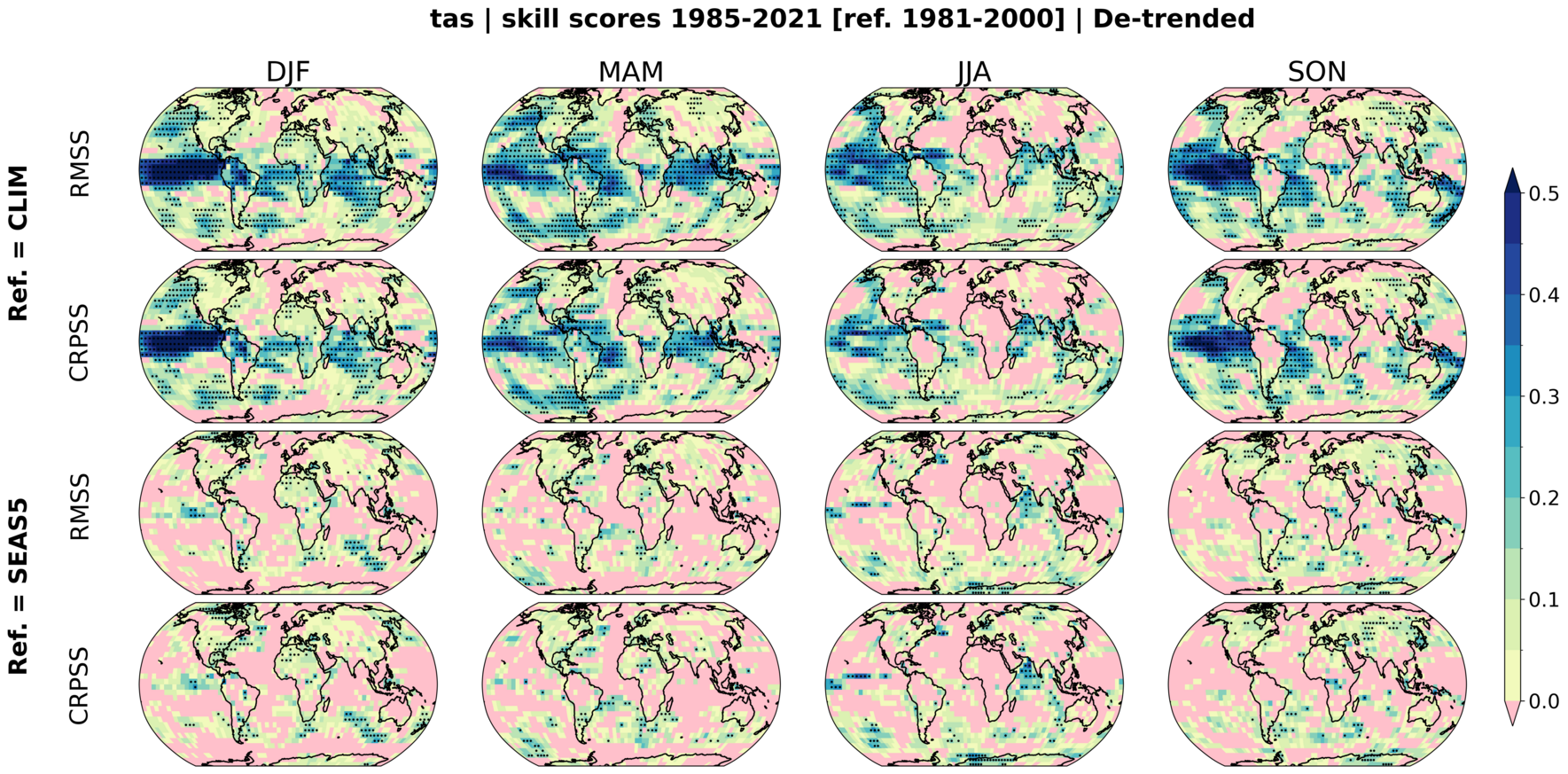}}
  \caption{Same as Figure \ref{f_glob_tas} but covering the period 1985-2021.
}\label{f_glob_skill_tas_det_extended}
\end{figure}

\begin{figure}[!h]
\centerline{\includegraphics[width=39pc]{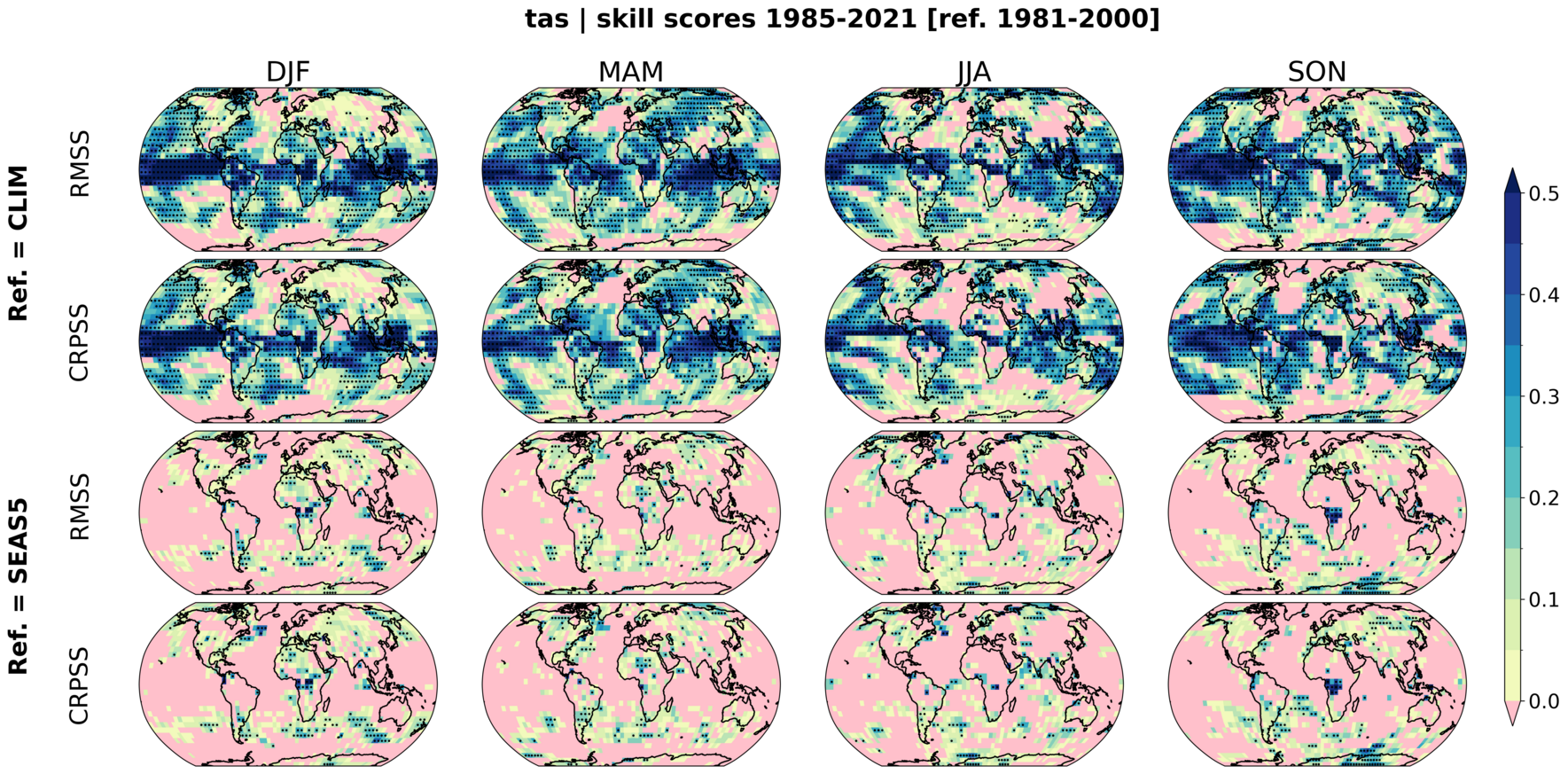}}
  \caption{Same as Figure \ref{f_glob_tas} but covering the period 1985-2021 and without de-trending.}\label{f_glob_skill_tas_nodet_extended}
\end{figure}

\begin{figure}[!h]
 \centerline{\includegraphics[width=39pc]{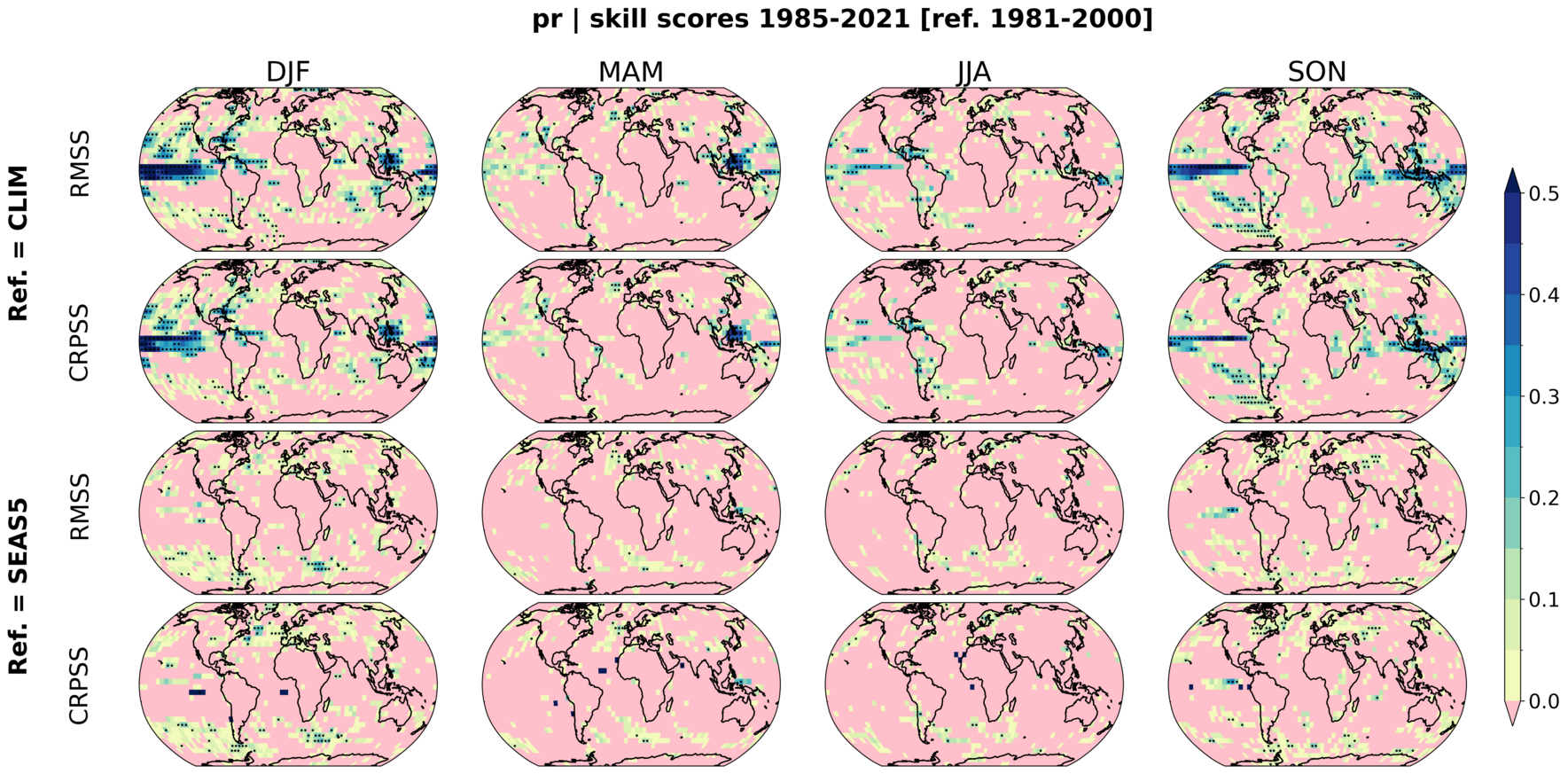}}
  \caption{Same as Figure \ref{f_glob_tas} but for precipitation and covering the 1985-2021 period.
}\label{f_glob_skill_pr_nodet_extended}
\end{figure}

% EU PLOTS

\begin{figure}[!h]
 \centerline{\includegraphics[width=39pc]{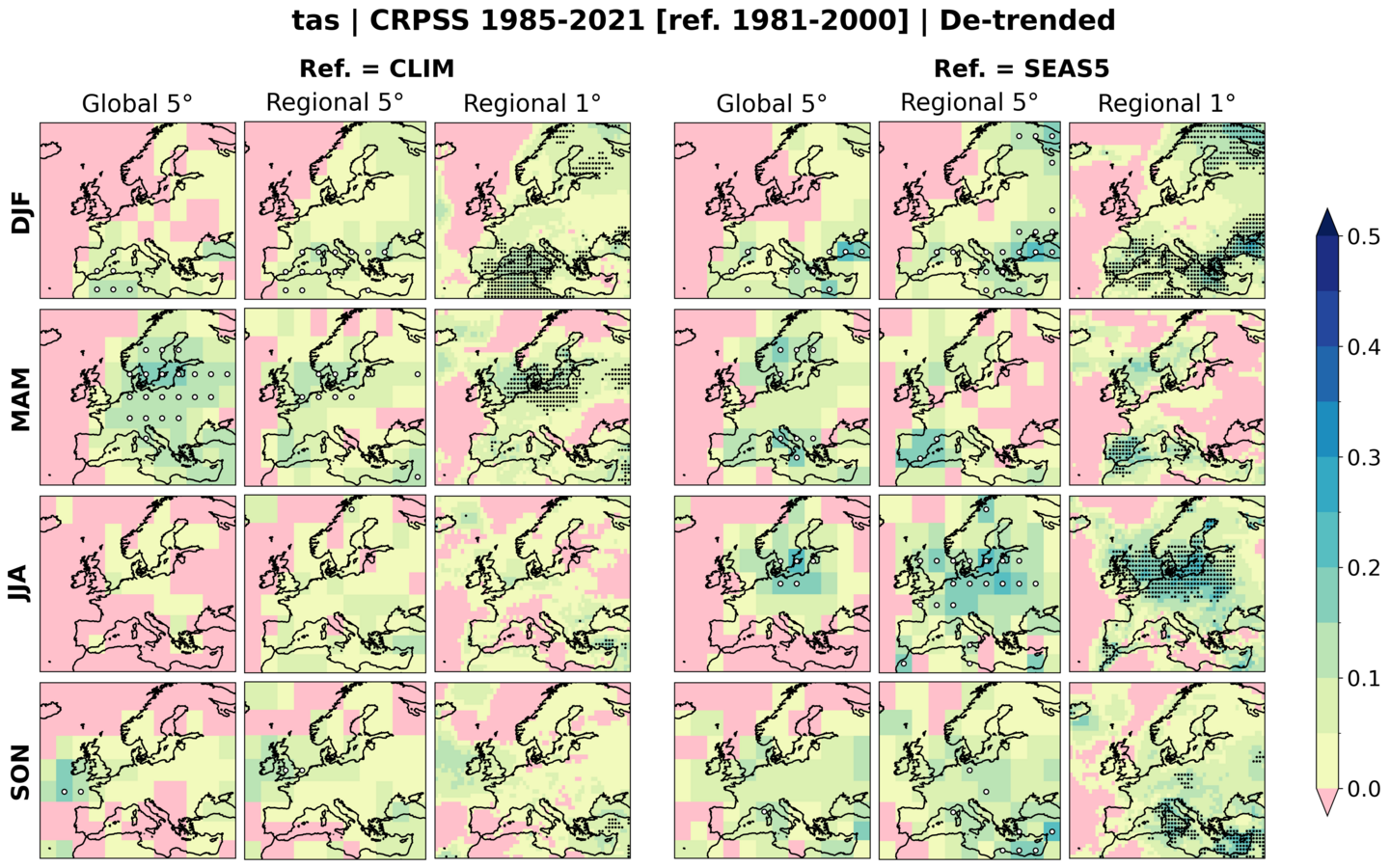}}
  \caption{Same as Figure \ref{f_eu_tas} but covering the 1985-2021 period. 
}\label{f_eu_skill_tas_det_extended}
\end{figure}

\begin{figure}[!h]
 \centerline{\includegraphics[width=39pc]{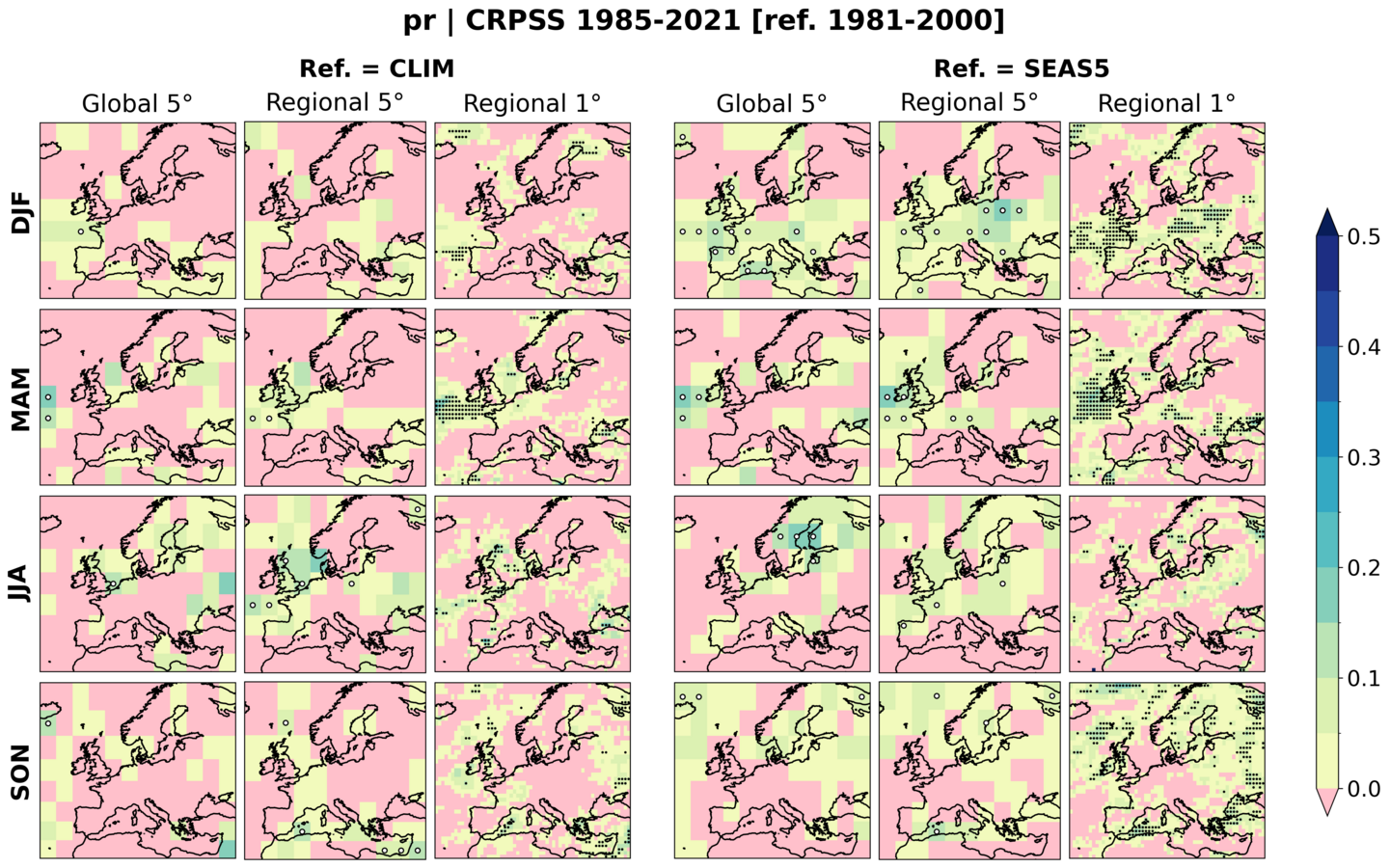}}
  \caption{Same as Figure \ref{f_eu_pr} but covering the 1985-2021 period. 
}\label{f_glob_skill_pr_nodet_extended}
\end{figure}

\clearpage
\appendix[B]\label{Appx_b}

\appendixtitle{Architecture details}

\subsection*{Hyperparemters configuration}

\begin{table}[h!]
\caption{Neural Network Architecture and Training Hyperparameters}\label{t1}
\begin{center}
\begin{tabular}{llr}
\topline
Hyperparameter & Description & Value \\
\midline
residual\_blocks\_decoder & Number of filters per residual block & [32, 32, 32, 32] \\
z\_size & Dimension of latent space & 128 \\
x\_encoder\_patch\_size & Input patch size for x encoder & 1 \\
x\_encoder\_embed\_dim & Embedding dimension for x encoder & 128 \\
x\_encoder\_n\_heads & Number of attention heads for x encoder & 1 \\
x\_encoder\_n\_layers & Number of transformer layers for x encoder & 8 \\
y\_encoder\_patch\_size & Input patch size for y encoder & 1 \\
y\_encoder\_embed\_dim & Embedding dimension for y encoder & 128 \\
y\_encoder\_n\_heads & Number of attention heads for y encoder & 1 \\
y\_encoder\_n\_layers & Number of transformer layers for y encoder & 8 \\
batch\_size & Training batch size & 32 \\
epochs & Number of training epochs & 100 \\
learning\_rate & Initial learning rate & 5e-5 \\
weight\_decay & L2 regularization parameter & 0.01 \\
loss\_lambda & Loss weighting parameter & 1 \\
loss\_alpha & Loss confidence parameter & 0.95 \\
\botline
\end{tabular}
\end{center}
\end{table}

\end{document}